%% file: manuscript_tg_GEM_uRWELL.tex
\documentclass[3p,times]{elsarticle}
\usepackage[breaklinks,hidelinks,colorlinks]{hyperref}

\usepackage{mathtools}
\usepackage{enumitem}
\usepackage{natbib}
\usepackage{soul}
\usepackage[dvipsnames]{xcolor}
\makeatletter
\renewcommand{\fnum@figure}{Fig. \thefigure}
\makeatother
\usepackage{lineno}

\begin{document}
\begin{frontmatter}
%
%
%
\title{Development of Thin-Gap GEM-\texorpdfstring{$\muup$}{}RWELL Hybrid Detectors}
%
%
\author[1]{Kondo Gnanvo\texorpdfstring{\corref{cor1}}{}} 
\texorpdfstring{\ead{kagnanvo@jlab.org}}{}
\cortext[cor1]{Corresponding author}
\author[1]{Xinzhan Bai}
\author[1]{Brian Kross}
\author[2]{Minh Dao}
\author[1]{Seung Joon Lee}
\author[2]{Nilanga Liyanage}
\author[2]{Huong Nguyen}
\author[4]{Matt Posik}
\author[5]{Nikolai Smirnov}
\author[1]{Sourav Tarafdar}
\author[1]{Andrew Weisenberger}
\address[1]{Thomas Jefferson National Accelerator Facility, Newport News, VA 23606, USA}
\address[2]{University of Virginia, Department Of Physics, Charlottesville VA 22903, USA}
\address[4]{Temple University, Philadelphia, PA 23606, USA}
\address[5]{Yale University, Physics Department, New Haven, CT 06520, USA}
%
\begin{abstract}
Micro Pattern Gaseous Detectors (MPGDs) are used for tracking in High Energy Physics and Nuclear  Physics because of their large area, excellent spatial resolution capabilities and low cost. However, for high energy charged particles impacting at a large angle with respect to the axis perpendicular to detector plane, the spatial resolution degrades significantly because of the long trail of ionization charges produced in clusters all along the track in the drift region of the detector. The long ionization charge trail results in registering hits from large number of strips in the readout plane which makes it challenging to precisely reconstruct the particle's position using simple center of gravity algorithm. As a result, the larger the drift gap, the more severe the deterioration of spatial resolution for inclined tracks. For the same reason, the position resolution is also severely degraded in a large magnetic field,  where the Lorentz E $\times$ B effect causes the ionization charges to follow a curved and longer path in the detector gas volume. 
In this paper, we  report on the development of thin-gap MPGDs as a way to maintain excellent spatial resolution capabilities of MPGD detectors over a wide angular range of incoming particles.  In a thin-gap MPGD, the thickness of the gas volume in the drift region is reduced from typically $\sim$3~mm to $\sim$1~mm or less. We present preliminary test beam results demonstrating  the improvement in spatial resolution from $\sim$400~$\muup$m with a standard 3~mm gap $\muup$RWELL prototype to $\sim$140~$\muup$m with a double amplification GEM-$\muup$RWELL thin-gap hybrid detector. We also discuss the impact of a thin-gap drift volume on other aspects of the performance of MPGD technologies such as the efficiency and  detector stability.
\end{abstract}
\end{frontmatter}
%
%
%
%
\input{01_introduction}
\input{02_prototypes}
\input{03_testbeamsetup}
\input{04_hv_scan}
\input{05_angle_scan}
\input{06_conclusion}
%
\bibliographystyle{elsarticle-num}
\bibliography{07_ref.bib}
\end{document}

%% file: 01_Introduction.tex
\section{Introduction}
In the barrel region of the central detector of a collider experiment, such as the ePIC detector in upcoming Electron Ion Collider (EIC) facility, large area micro pattern gaseous detectors (MPGDs) are a cost effective solution  to provide precise tracking capability over a wide pseudorapidity range  inside a strong  magnetic field. Standard MPGD detectors require an uniform drift gap of the order of 3~mm or larger, where the ionization due to incident high energy particles takes place. The electrons from the primary and secondary ionization follow the field lines to drift vertically toward the amplification structure. The electronic signal induced on the anode readout plane by the  charge cluster cloud from the amplification process is a projection of the charge clusters produced along the ionization trail left by the high energy particle in the drift gap. When the incident particle is perpendicular to  or impinges on the detector at a narrow angle ($\leq 5^{\circ}$) with respect to the vertical axis to the detector plane, the electronic signal is induced only on a small cluster of neighboring strips of the order of 3 or 4 on average. In this case, the center of gravity (COG) algorithm is applied by weighting the position of each strip in the cluster according to its ADC value, allowing for the reconstruction of the particle position with high precision, typically one order of magnitude better than the pitch size of the readout strips. However, when particles hit the detector at a larger angle, the induced signal affects a much larger area of the strip readout plane because of the longer ionization trail and the signal is distributed over a larger number of strips. 
\begin{figure}[!ht]
\centering
\includegraphics[width=0.95\columnwidth,trim={0pt 0mm 0pt 75mm},clip]{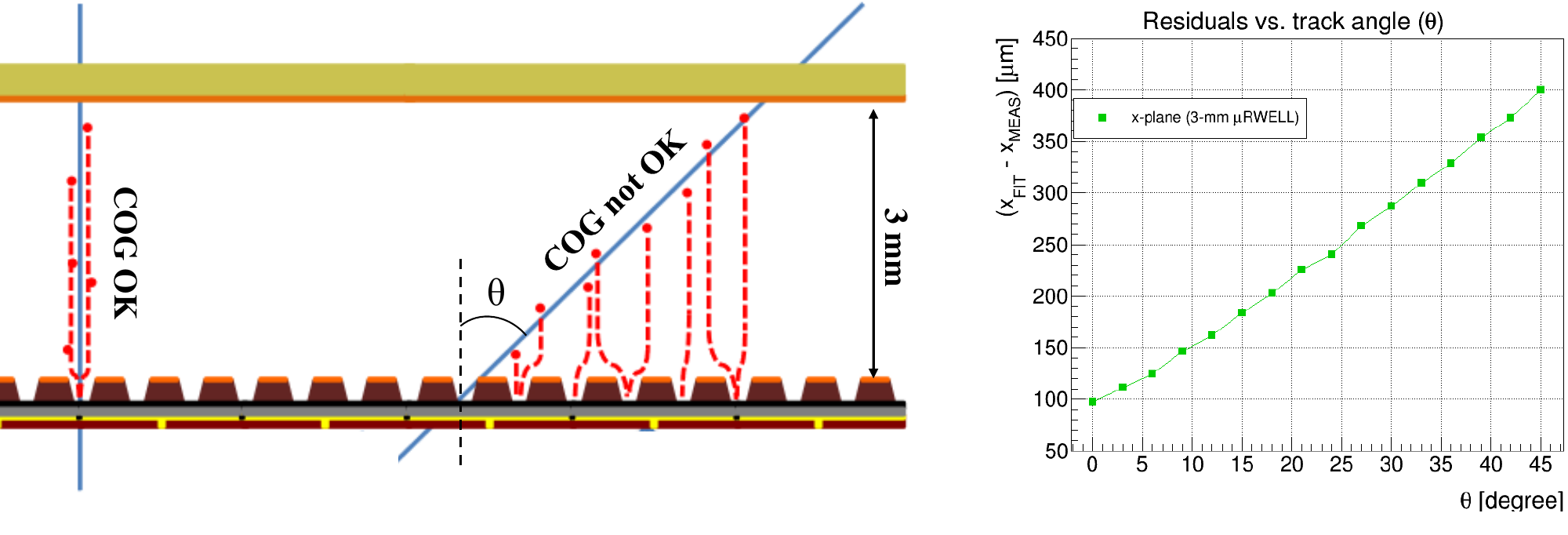}
\caption{\label{fig:stdgap_vs_resol} \textit{(Left:)} Cross sectional pictorial representation of the ionization trail of a particle traversing a standard MPGD detector with a 3-mm gap gas volume in the ionization region, at an angle of 0$^\circ$ and 45$^\circ$. \textit{(Right:)} Plot of the spatial resolution using center of gravity (COG) method as a function of the impact angle of the 120 GeV proton beam on a 3-mm gap $\muup$RWELL detector from the beam test data described in section ~\ref{sec:testbeam}.}
\end{figure}
In this case, the COG algorithm is no longer effective in determining the particle position with good accuracy. This is illustrated in the cross sectional representation of a  3~mm gap $\muup$RWELL detector on the left side of Fig.~\ref{fig:stdgap_vs_resol}. The plot on the right shows the degradation of the spatial resolution as a function of the impinging angle of  particles from test beam data conducted at the Fermilab Test Beam Facility (FTBF) in Summer 2023.\\
There are essentially two approaches to recover good spatial resolution performance for large angle tracks crossing MPGD trackers. The first, often referenced in the literature as $\muup$TPC (or mini-drift)~\cite{uTPC_Bencivenni_2021,uTPC_Azmoon_2016,uTPC_Iodice_2012}, involves operating the MPGD tracker like a TPC but with a significantly smaller drift length, typically $\sim$6 to 20 mm, which is about two orders of magnitude smaller than that of the standard TPC detector. Operating a MPGD tracker with such a small drift gap in $\muup$TPC mode results in a number of constraints and requirements that make the application as large area MPGD trackers in a real experimental environment  challenging. Past studies of MPGD trackers operating in $\muup$TPC mode\cite{uTPC_Bencivenni_2021,uTPC_Azmoon_2016,uTPC_Iodice_2012}  have shown that spatial resolution performance is worst for tracks perpendicular and almost perpendicular (within 0-15$^\circ$) to detector plane.  In the end cap region of collider experiments where a significant number of tracks impinge the detector plane almost perpendicularly at higher pseudo-rapidity, operating MPGD tracker in $\muup$TPC mode would constitute a poor choice. Additionally, $\muup$TPC mode works effectively with fast electronics, which would increase the cost, and slower gases, which would degrade the detector timing performance. The second approach, which is under consideration in this paper, is the development of thin-gap MPGDs. Unlike the $\muup$TPC-MPGDs, in  thin-gap MPGDs, the  drift  gap is reduced from $\sim$3~mm to $\sim$1~mm or less to minimize the spatial extension of the ionization trail of the incoming particle, thereby minimizing its negative impact on the degradation of the spatial resolution at large angle tracks. A similar concept based on wire chamber~\cite{sTGC}, the small strip thin gap chamber (sTGC) was developed for the CERN ATLAS Muon system upgrade. The concept of thin-gap MPGD also comes with its own set of challenges, particularly the efficiency drop that derives from the smaller number of primary and total ionization charges produced in the drift region due to the reduced gap. Additionally, the technical challenge of maintaining a uniform gap in the gas volume in the drift region between the cathode and the amplification structure should not be ignored. \\
In the current paper, we present the novel concept of a thin-gap  GEM-$\muup$RWELL hybrid detector, consisting of a stack of a cathode foil, a GEM foil~\cite{SAULI:1997nim} for electron pre-amplification, a $\muup$RWELL device~\cite{Bencivenni:2015} for a second-stage amplification, and a high performance three-layer stack of capacitive-sharing X--Y strip readout\cite{capaSh_urwell2022} as the anode PCB. In section~\ref{sec:tgrwell}, we introduce the concept of a thin-gap  GEM-$\muup$RWELL hybrid detector and describe the design and construction of prototypes. In section~\ref{sec:testbeam}, we present the studies of two thin-gap  GEM-$\muup$RWELL prototypes in a high energy proton beam at the FTBF. In sections~\ref{sec:hv_scan}~and~\ref{sec:angle_scan}, we present the performance of the prototypes during high voltage scans of the various amplification layers including the study of spatial resolution and detector efficiency as a function of the angle of the impinging  particle. Finally, in section~\ref{sec:conclusion}, we provide the initial conclusions on the thin-gap GEM-$\muup$RWELL hybrid concept and discuss the perspectives of these novel detectors in future experiments.

%% file: 02_prototypes.tex
\section{Development of the thin-gap GEM-\texorpdfstring{$\muup$}{}RWELL hybrid detectors.}
\subsection{Basic concept and operating principle of thin-gap GEM-\texorpdfstring{$\mu$}{}RWELL hybrid detectors.}
\label{sec:tgrwell}
The motivation for developing thin-gap GEM-$\muup$RWELL hybrid detectors is to improve the performance of the detectors in terms of efficiency, spatial and timing resolution over a large angular range of the impacting particles, as well as operation stability. Three distinct features are implemented in the thin-gap  GEM-$\muup$RWELL hybrid detector.
\begin{figure}[!ht]
\centering
\includegraphics[width=0.95\columnwidth,trim={0pt 55mm 0pt 35mm},clip]{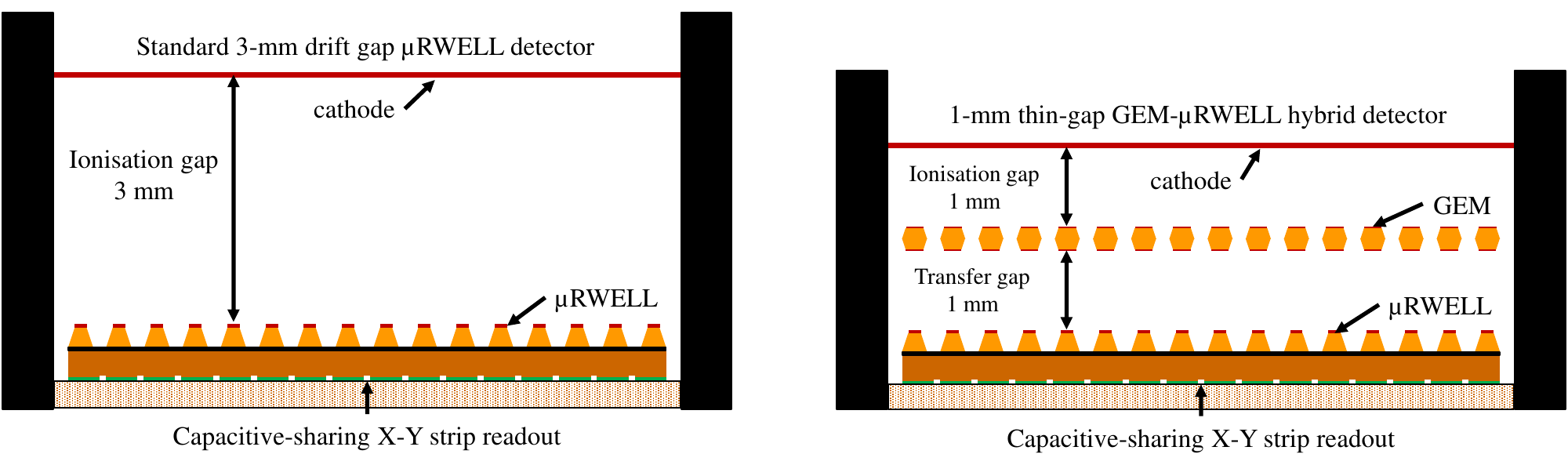}
\caption{\label{fig:std_vs_tg} \textit{(Left:)}  Cross sectional view of a standard 3-mm gap $\muup$RWELL detector;  \textit{(right:)} Cross sectional view of a 1-mm gap thin-gap GEM-$\muup$RWELL hybrid detector.}
\end{figure}
%
%
\begin{enumerate}[leftmargin=*]
\item \textbf{Thin drift gap:} The basic idea of a thin-gap GEM-$\muup$RWELL hybrid detector is to reduce the gap in the ionization region between the cathode and the  amplification layer from $\sim$3~mm or more, typically used in  standard MPGDs, to $\sim$1~mm or less as illustrated on the cartoon pictures of Fig.~\ref{fig:std_vs_tg}. A thinner ionization gap reduces the length of the ionization trail left by particles traversing the detector at large angles and subsequently limits the impact on the degradation of the spatial resolution.
\begin{figure}[!ht]
\centering
\includegraphics[width=0.95\columnwidth,trim={0pt 0mm 0pt 0mm},clip]{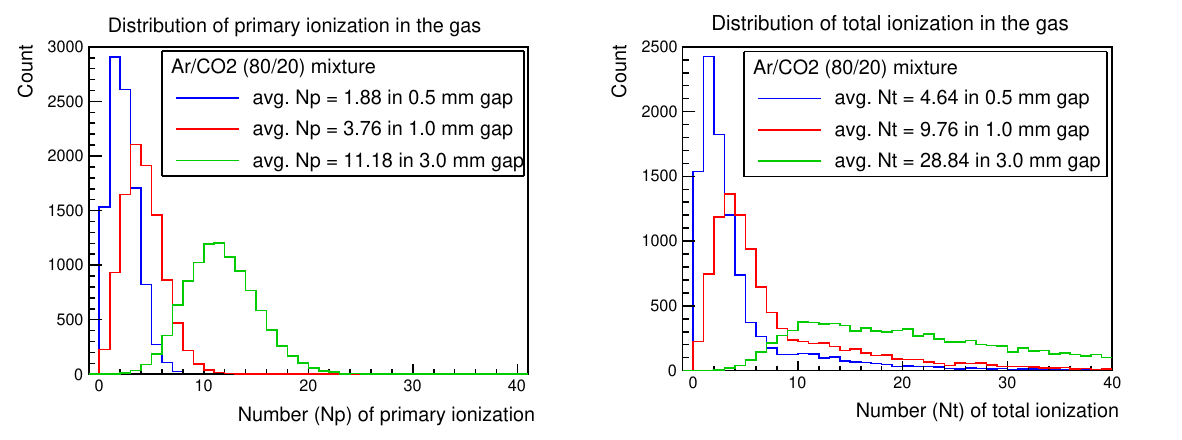}
\caption{\label{fig:ionization}  Distribution of number of primary ionization charges \textit{(left)}; and total ionization \textit{(right)} by a high energy particle traversing a gas length of  0.5~mm (green); 1~mm (blue), 3~mm Red), in Ar:CO2 8:20 gas mixture.
}
\end{figure}
The major drawback of thin-gap MPGDs with reduced number of ionization charges produced in the drift volume, is the drop of detector efficiency due to a smaller electronic signal amplitude. The Garfield++ simulation tool~\cite{Garfield} was used to study how the drift gap size impacts the number ionization charges for different gas mixtures. The drift volume was simulated as a box of gas whose thickness represents the drift gap size. The gas mixtures and ionization charges were simulated using Garfield's Magboltz~\cite{Magboltz} and HEED~\cite{SMIRNOV2005474} software packages. The plots in Fig.~\ref{fig:ionization} show the ionization charges produced in Ar:CO$_2$ (80:20) gas mixture by a minimum ionizing particle (MIP) for three gas volume defined by 0.5~mm, 1~mm and 3~mm gaps. The plots on the left show that the average number of primary ionization charges drops linearly with the gas thickness, from $\sim$11.18 for 3~mm gap to $\sim$3.78 for 1~mm gap, and $\sim$1.88 for 0.5~mm gap. Subsequently, on right of Fig.~\ref{fig:ionization}, the total ionization charge also drops by a factor of 3, from $\sim$28.8 to $\sim$9.8 on average, as the gap reduces from 3~mm gap to 1~mm. 
\item \textbf{Hybrid amplification:} The smaller ionization charge produced in the drift region of a thin-gap MPGD is in part compensated by increasing the gain of the amplification structure of the detector. In a single-amplification stage device such as a $\muup$RWELL, this is achieved by increasing the bias voltage (electric field) across the $\muup$RWELL structure, between the Cu electrode layer at the top and the diamond-like carbon (DLC) resistive layer at the bottom. However, the required voltage range for stable operation at full efficiency of single-amplification MPGD devices is usually very limited for standard (3-mm) gap detectors, especially when coupled with two-dimensional strip readout anode, where the signal from the ionization charge cluster is shared by two readout strip layers. As such, it is extremely challenging for single-amplification thin-gap MPGD like $\muup$RWELL to increase bias voltage to reach a detector gain three times higher as a way to compensate for the reduced ionization charges in the drift gap shown in the right of Fig.~\ref{fig:ionization}. A new approach using a double and hybrid amplification scheme is proposed in this paper to achieve large detector gain. In this scheme, the detector combines two stages of electron multiplication. The first is a GEM foil placed between the cathode and the $\muup$RWELL layer, which provides a pre-amplification gain on the order of 20 to 50. This is followed by the second amplification provided by the  $\muup$RWELL structure with an additional gain factor between 1,000 and 2,000.  In total, the detector can achieve a total gain ranging from 20,000 to 100,000. The concept of hybrid GEM-$\muup$RWELL was in the past suggested and tested by~\cite{SHEKHTMAN2019401} with standard 3 mm drift gap GEM-$\muup$RWELL prototype. The hybrid amplification structure ensures simultaneously a large gain to compensate for the deficit of ionization charges in the drift region of thin-gap MPGDs and the flexibility to fine tune the applied voltage in different regions of the detector to ensure stable operation. 
\item \textbf{Capacitive-sharing 2D readout:} The concept of capacitive-sharing 2D-strip readout with MPGDs is described in full detail in~\cite{capaSh_urwell2022, capaSh_gemxyu2023}. This technology is a critical component of thin-gap MPGDs that allows to achieve excellent spatial resolution performance. A diagram of a cross-sectional view of the readout concept is shown in Fig.~\ref{fig:capash-grwell}.   Indeed, the narrow drift gap and reduced number of ionization charges of a thin-gap MPGD result in a smaller lateral extension of the ionization charge cloud after amplification. A thin-gap GEM-$\muup$RWELL with standard strip readout layer, will require higher granularity with strip segmentation pitch of the order of 100~$\muup$m to achieve a spatial resolution performance.
\begin{figure}[!ht]
\centering
\includegraphics[width=0.9\columnwidth,trim={0pt 0mm 0pt 25mm},clip]{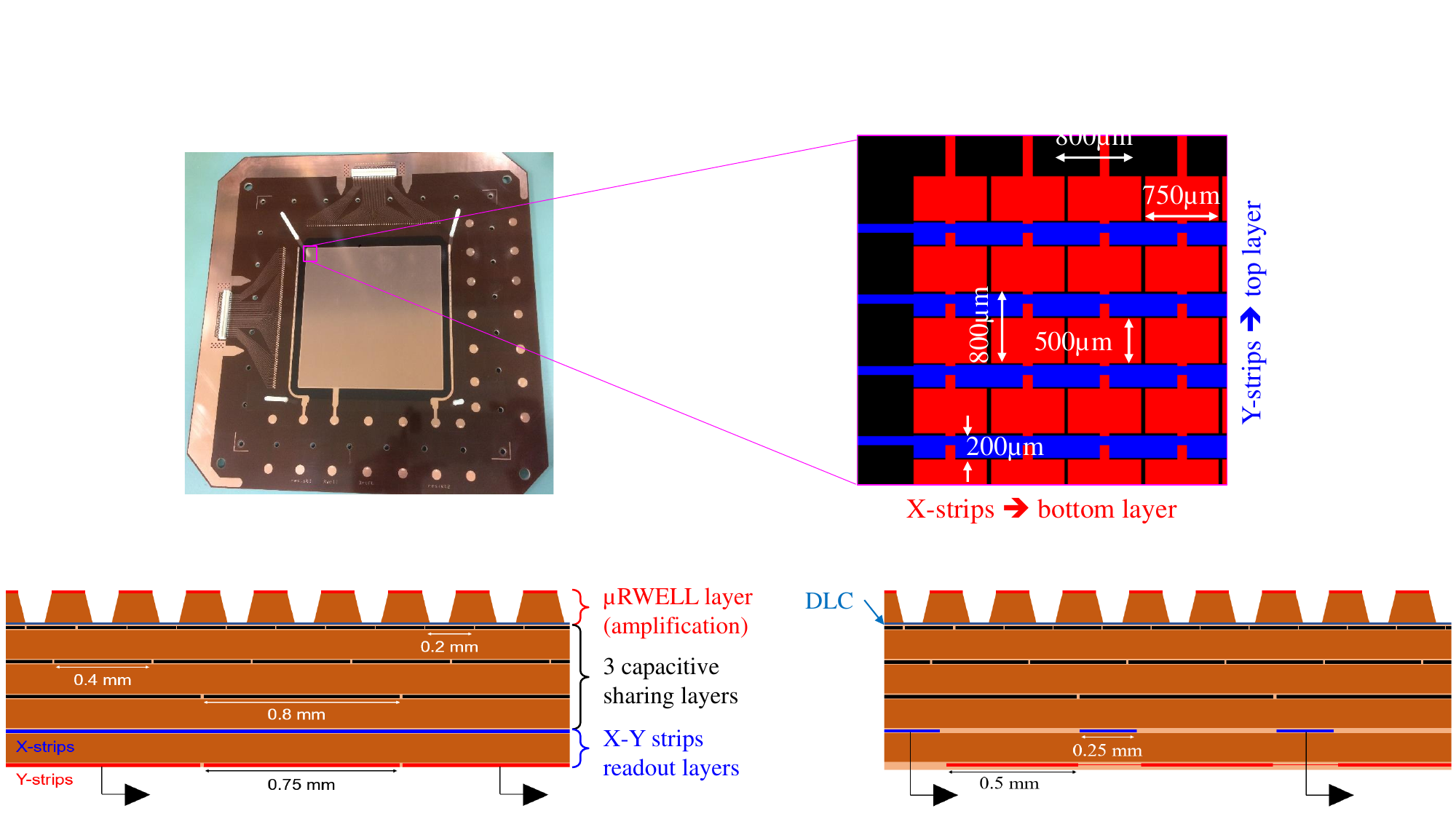}
\caption{\label{fig:capash-grwell} \textit{(Top left:)} Picture of a $\muup$RWELL PCB using capacitive-sharing X--Y strip readout; \textit{(top right:)} zoomed view of the X--Y strip readout;  \textit{(bottom left:)} Cross-sectional view of the X--strip layer;  \textit{(bottom right:)} Y--strip layers of capacitive-sharing readout.}
\end{figure}
Capacitive-sharing readout structures with strip pitch as large as 1.6~mm, could provide spatial resolution better than 100~$\muup$m, comparable to the performance obtained with a triple-GEM with 400~$\muup$m pitch COMPASS X-Y strip readout~\cite{COMPASS:2002}, resulting to a significant reduction of the number of electronic channels to read out. 
\end{enumerate}
%
%
\subsection{Assembly of thin-gap GEM-\texorpdfstring{$\mu$}{}RWELL prototypes.}
\label{sec:jlab_protos}
\begin{figure}[!ht]
\centering
\includegraphics[width=0.995\columnwidth,trim={0pt 30mm 0pt 0mm},clip]{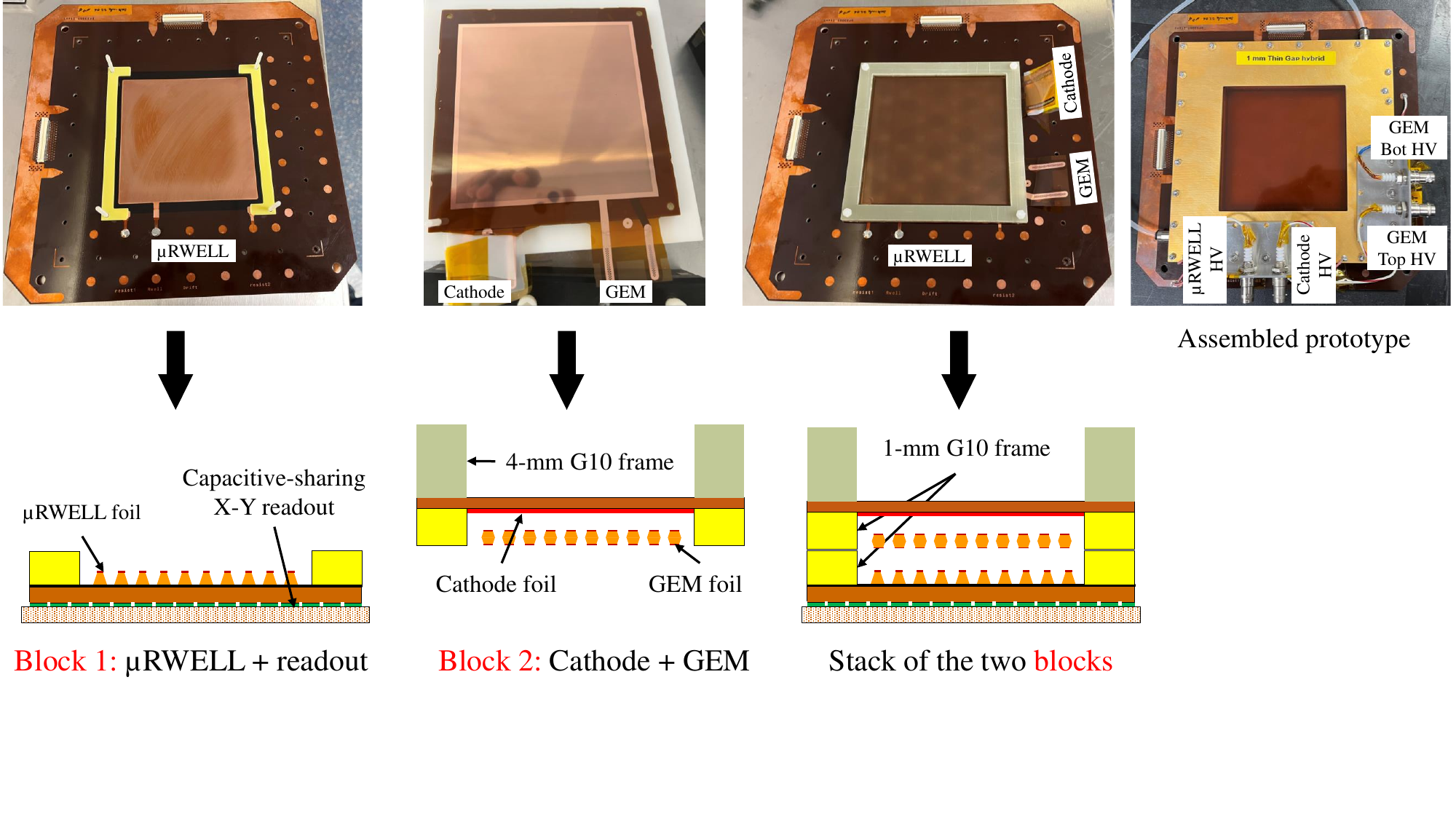}
\caption{\label{fig:jlab_protos} Assembly in clean room of 1-mm thin-gap GEM-$\muup$RWELL hybrid prototype - \textit{(left:)} $\muup$RWELL foil with capacitive-sharing X--Y strip readout PCB; \textit{(center left:)} framed cathode and GEM foils glued to a 4-mm thick G10 outer frame as a block; \textit{(center right:)} Stack of the $\muup$RWELL PCB and the cathode-GEM block;  \textit{(right:)} Final prototype with four SHV connectors to the cathode, top and bottom GEM and  $\muup$RWELL electrode.}
\end{figure}

Two  small (10 cm $\times$ 10 cm) thin-gap GEM-$\muup$RWELL prototypes named proto-I and proto-II, were fabricated in the detector lab of the Radiation Detector and Imaging Group (RD\&I Group) at Jefferson Lab and tested in a high energy proton beam at the FTBF. The GEM foils and the multi-layer PCB combining the $\muup$RWELL amplification structure with the capacitive-sharing X--Y strip readout planes~\cite{capaSh_urwell2022}, were procured at the CERN Micro Pattern Technology (MPT) Workshop. A picture of the $\muup$RWELL PCB is shown in the top left of Fig.~\ref{fig:capash-grwell}.  A zoomed-in view of the X--Y strip readout layer of the PCB is seen in the top-right diagram of Fig.~\ref{fig:capash-grwell} and the cross sectional views illustrating $\muup$RWELL amplification device, the stack of the three-layer capacitive-sharing structure and the two-layer X--Y strip readout are shown at the bottom -- along the x-direction (left) and the y-direction (right). The prototypes were assembled in the MPGD clean room at Jefferson Lab. Both prototypes are based on the design shown in Fig.~\ref{fig:std_vs_tg}, except for the thickness of the gas volume in the drift region which is 1~mm for proto-I and 0.5~mm for proto-II. The pictures and diagrams in Fig.~\ref{fig:jlab_protos} illustrate the assembly steps carried out in the clean room for proto-I.
On the left, the $\muup$RWELL with the capacitive-sharing readout PCB ($\muup$RWELL + readout block) as received from CERN is shown on the top picture. 
The central panel is a picture of cathode+GEM consisting of a stack of 4 mm thick G10 frame (green block in the diagram below) glued to the top side of cathode foil. The cathode foil is made of a 5~$\muup$m Cu laminated on a 50~$\muup$m Kapton foil as shown by red and brown lines respectively in the bottom cross section view. The GEM foil is stretched and glued to a 1~mm G10 frame (shown by yellow blocks in bottom figure) which is glued to the bottom of the cathode foil. The 1~mm G10 frame defines the ionization gap between the cathode and GEM foil. The cathode-GEM block is then stacked on top of the $\muup$RWELL PCB separated by 1-mm thick yellow G10 spacers shown in the right picture of  Fig.~\ref{fig:jlab_protos}. Finally a gas enclosure with top cover lid having a Kapton foil window in the active area closes the double amplification stack with Buna-N O-rings and a set of screws and bolts to ensure gas tightness. Four SHV connectors, (picture in right of Fig.~\ref{fig:jlab_protos}), are mounted on the top of the detector to supply the voltage to the cathode, the top and bottom GEM and the $\muup$RWELL electrodes.

%% file: 03_testbeamsetup.tex
\section{Thin-gap GEM-\texorpdfstring{$\muup$}{}RWELL prototypes in beam test at Fermilab}
\label{sec:testbeam} 
The two prototypes were installed in a tracking telescope and tested in the 120 GeV proton beam at Fermilab  during the summer 2023. The test beam setup and experimental conditions are described in the following subsections
%
%
\subsection{Fermilab test beam setup}
\begin{figure}[!ht]
\centering
\includegraphics[width=0.995\columnwidth,trim={0pt 0mm 0pt 0mm},clip]{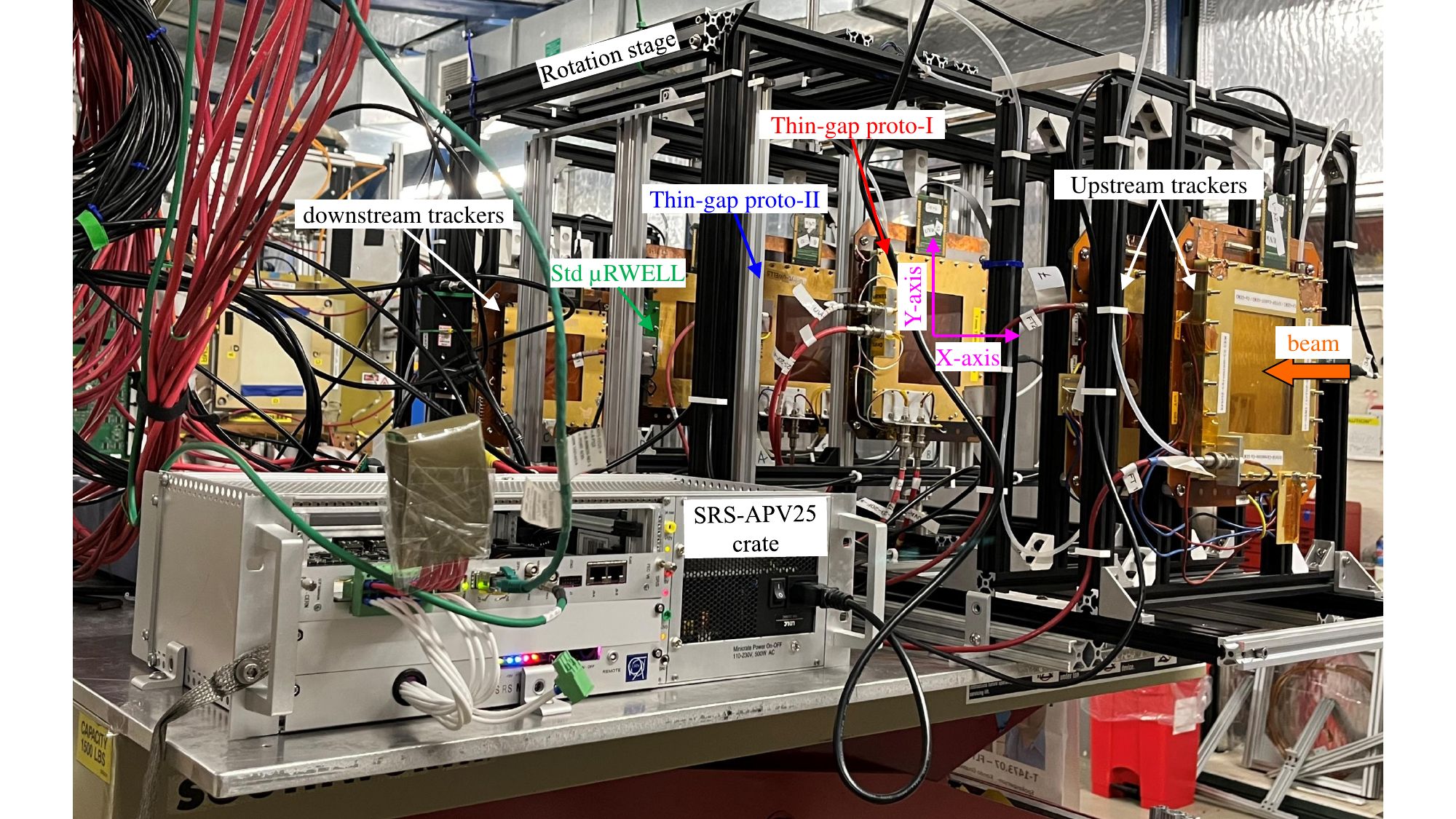}
\includegraphics[width=0.9\columnwidth,trim={0pt 0mm 0pt 105mm},clip]{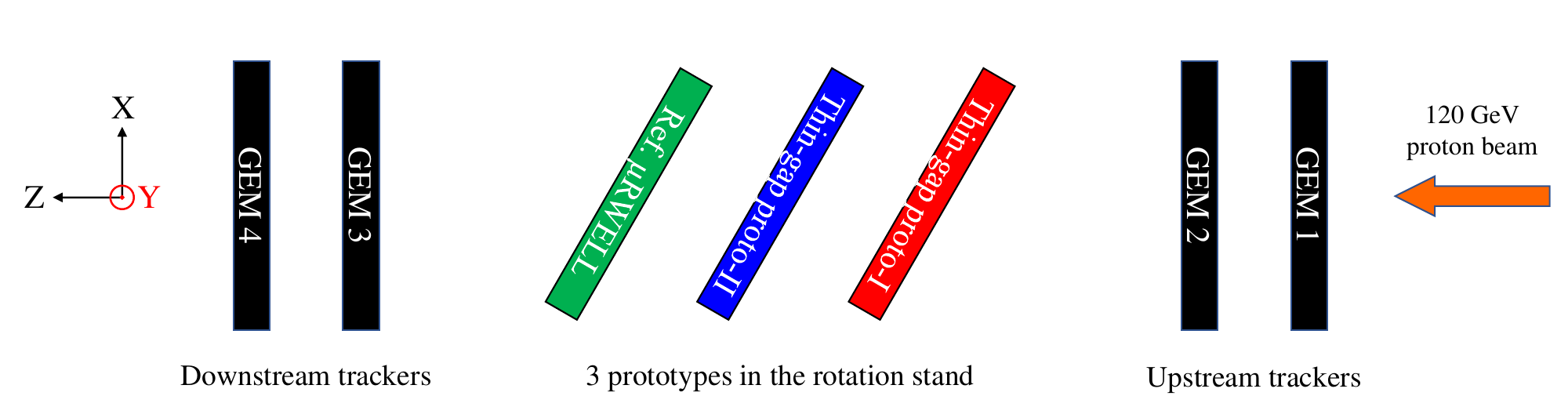}

\caption{\label{fig:fnalsetup} Setup of the two thin-gap GEM-$\muup$RWELL hybrids and a standard 3-mm gap $\muup$RWELL prototypes in the rotation stage with two sets of two GEM reference trackers upstream and downstream in the MT62.b stage of the FTBF. The SRS-APV readout  crate is shown on the left}
\end{figure}

The tracking telescope, shown in the picture of Fig.~\ref{fig:fnalsetup} was set up in the MT6.2b area at the FTBF for the characterization of the thin-gap prototypes. It consisted of two pairs of standard 10~cm~$\times$~10~cm triple-GEM detectors, mounted upstream and downstream of a motorized plane rotation stage described in section~\ref{subsec:rotstand} where the two thin-gap GEM-$\muup$RWELL prototypes, together with a 3-mm drift gap $\muup$RWELL used as a reference detector were mounted. 
The characteristics of the three prototypes under study and the four reference GEM trackers  are listed in Table~\ref{table:jlab_tab1}. 
\begin{table} [!ht]
\begin{center}
\begin{tabular}{|p{5.2cm}|p{2.cm}|p{2.cm}|p{5.5cm}| }
\hline
\multicolumn{4}{|c|}{MPGDs in FTBF test beam setup} \\
\hline
Prototypes / trackers & Drift gap  & Transfer gap & 2D-strip readout structure / pitch \\
\hline
2 $\times$ triple-GEM trackers upstream & 3.0 mm & 2.0 mm & COMPASS GEM X--Y strip / 0.4~mm\\\hline
Thin-gap GEM-$\muup$RWELL proto-I   & 1.0 mm  & 1.0 mm & CapaSh X--Y strip / 0.8~mm\\
\hline
Thin-gap GEM-$\muup$RWELL proto-II  & 0.5~mm & 1.0 mm & CapaSh X--Y strip /  0.8~mm\\
\hline
standard $\muup$RWELL - reference & 3.0 mm & N/A & CapaSh X--Y strip, pitch / 0.8 mm\\
\hline
2 $\times$ triple-GEM trackers downstream & 3.0 mm & 2.0 mm & COMPASS GEM X--Y strip / 0.4~mm\\
\hline
\end{tabular}
\caption{\label{table:jlab_tab1}{Thin-gap GEM-$\muup$RWELL prototypes and GEM trackers in FTBF test beam setup}}
\end{center}
\end{table}
%
%
\subsection{Rotation stage for angle scan measurements}
\label{subsec:rotstand}
The rotation stage is designed to accommodate up to three prototypes at the same time and is motorized to rotate the vertical axis of the X--Y plane of the prototypes allowing the incoming particles of the proton beam to impact the detector at angles ranging from 0 to 45 degrees.  The stage is a 70 cm $\times$ 60 cm $\times$ 70 cm  support constructed using 80/20 aluminum extrusion rails. The three mounting 80/20 aluminum frames were connected by a linkage that makes the same rotation angle for all three mounting frames. 
\begin{figure}[!ht]
\centering
\includegraphics[width=0.975\columnwidth,trim={0pt 30mm 0pt 20mm},clip]{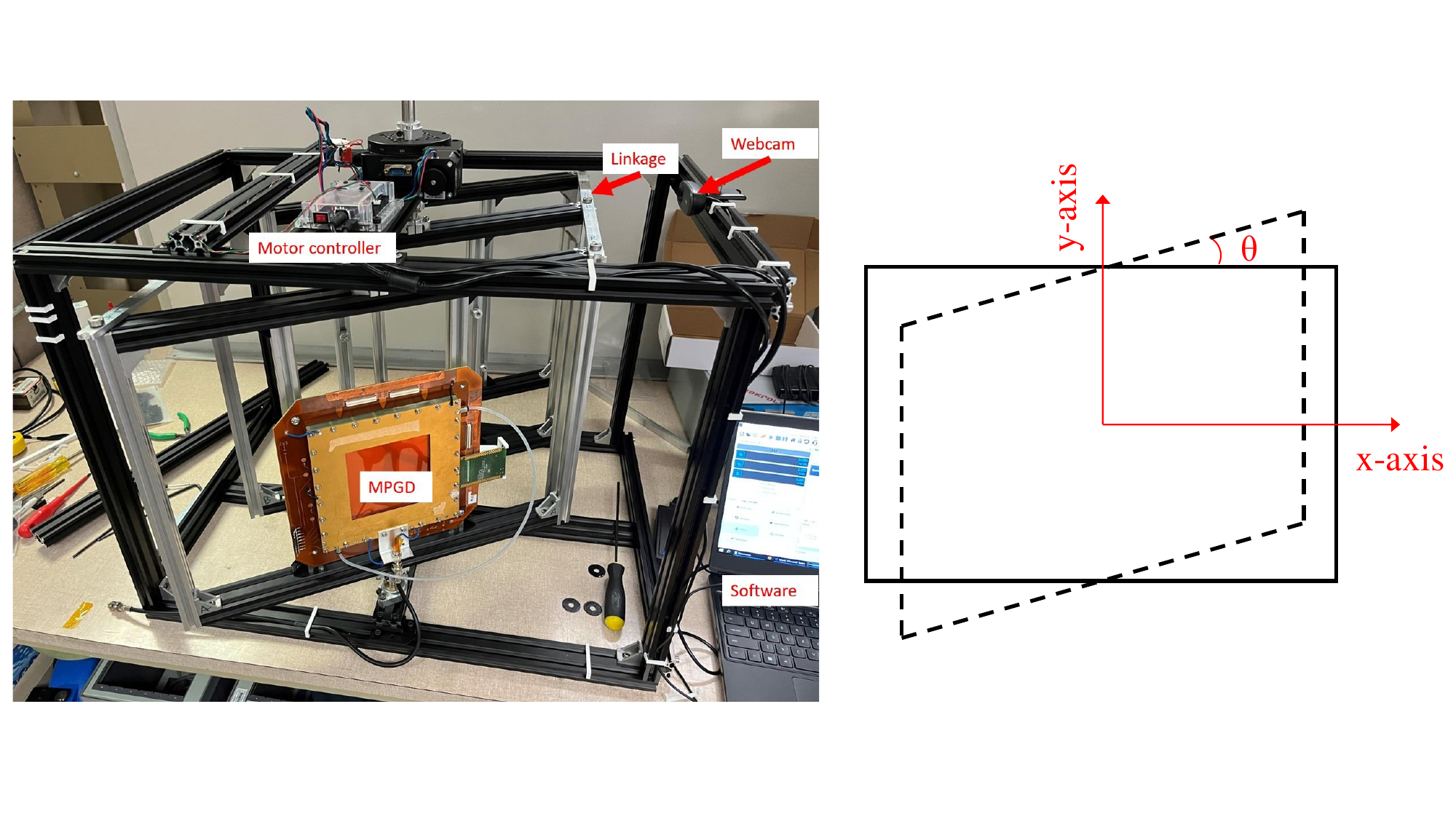}
\caption{\label{fig:rotstand} \textit{(Left:)} Picture of the mechanical frame of the rotation stand with the motor controller and a webcam for online monitoring of the rotation. One of the MPGD prototypes is shown mounted for illustration; \textit{(right:)} schematic of the plane rotation around the y-axis.}
\end{figure}
The stepper motor driven rotation stage has worm/spur gears with a 180:1 gear ratio. The stepper motor (1.8\textdegree  per step) was controlled with a GRBL open source motion controller with a resolution of  0.01 degree / step and the estimated backlash is 0.1 degree. The controller was connected to a PC via a USB port and run by an open source software. For the experiment at FTBF, an Ethernet based USB extender was used to control the stage from the control room. A USB webcam was attached to the rotation stage to monitor the rotation stage from the control room.
%
%
\subsection{Gas mixing system}
\label{subsec:gasmixing}
\begin{figure}[!ht]
\centering
\includegraphics[width=0.95\columnwidth]{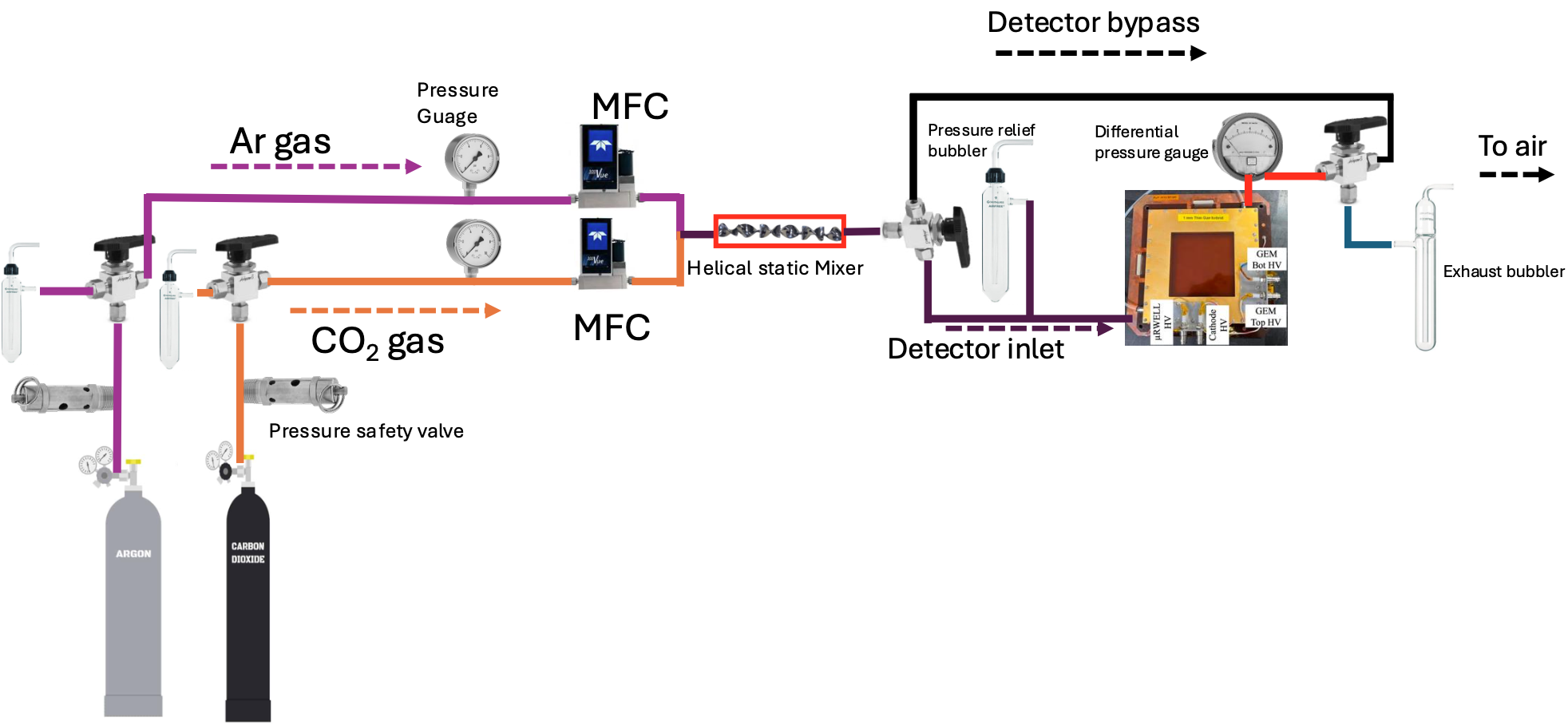}
\caption{\label{fig:gas_mix} Schematics of gas mixing unit used in Fermilab test beam campaign for the thin gap prototypes.}
\end{figure}
All the four reference GEM trackers, the 3-mm gap $\muup$RWELL reference detector and the two thin-gap GEM-$\muup$RWELL prototypes under study were operated with a Ar:CO$_2$ (80:20) gas mixture. 
The gas supply was controlled through a gas mixing unit capable of mixing two different gases at various proportion. A schematic of the gas mixing unit is shown in Fig.~\ref{fig:gas_mix}. The mixing ratio is determined by the flow rate controlled by mass flow controllers labeled as \textbf{MFC} in Fig.~\ref{fig:gas_mix}. The output pressure of the gas bottles was maintained below 10 psi (69 kPa). To avoid accidental over pressure inside the gas mixing unit components the inlets to the mass flow controllers were fitted with pressure safety valves. The inlet to the detector gas line from the gas mixing unit has a pressure-relief bubbler to limit the pressure inside the detector. The differential pressure in the detector was monitored by using differential pressure gauge. Throughout the experiment the detector differential pressure was kept below 50 Pa which means the pressure inside the detector was almost the same as atmospheric pressure. The gas mixing unit was capable of mixing two gases within 1$\%$ accuracy. During the operation of the detectors the total flow rate from the mixing unit was maintained at 100 sccm.  The gas from the outlet of the detectors was released into the atmosphere.  
%
%
\subsection{SRS-APV25 readout electronics and DAQ system}
\label{subsec:srs}
All detectors were read out with the APV25-based~\cite{apv25} Scalable Readout System (SRS)~\cite{srs2011} developed by the CERN RD51 collaboration \cite{rd51coll}. The APV25-SRS front-end (FE) hybrids on the detectors were connected via HDMI cables to the SRS-ADC cards of the SRS crate \cite{srs2011} as shown in the bottom left of  Fig.~\ref{fig:fnalsetup}. In the FTBF setup the data acquisition has a sampling rate of 50 ns and nine APV25 time samples corresponding to an acquisition window of 450 ns per readout channel. A coincidence signal formed by a set of three trigger counters provided as part of the equipment of the MT6.2b area was used to externally trigger the SRS crate. 
\begin{figure}[!ht]
\centering
\includegraphics[width=0.9\columnwidth,trim={0pt 35mm 0pt 25mm},clip]{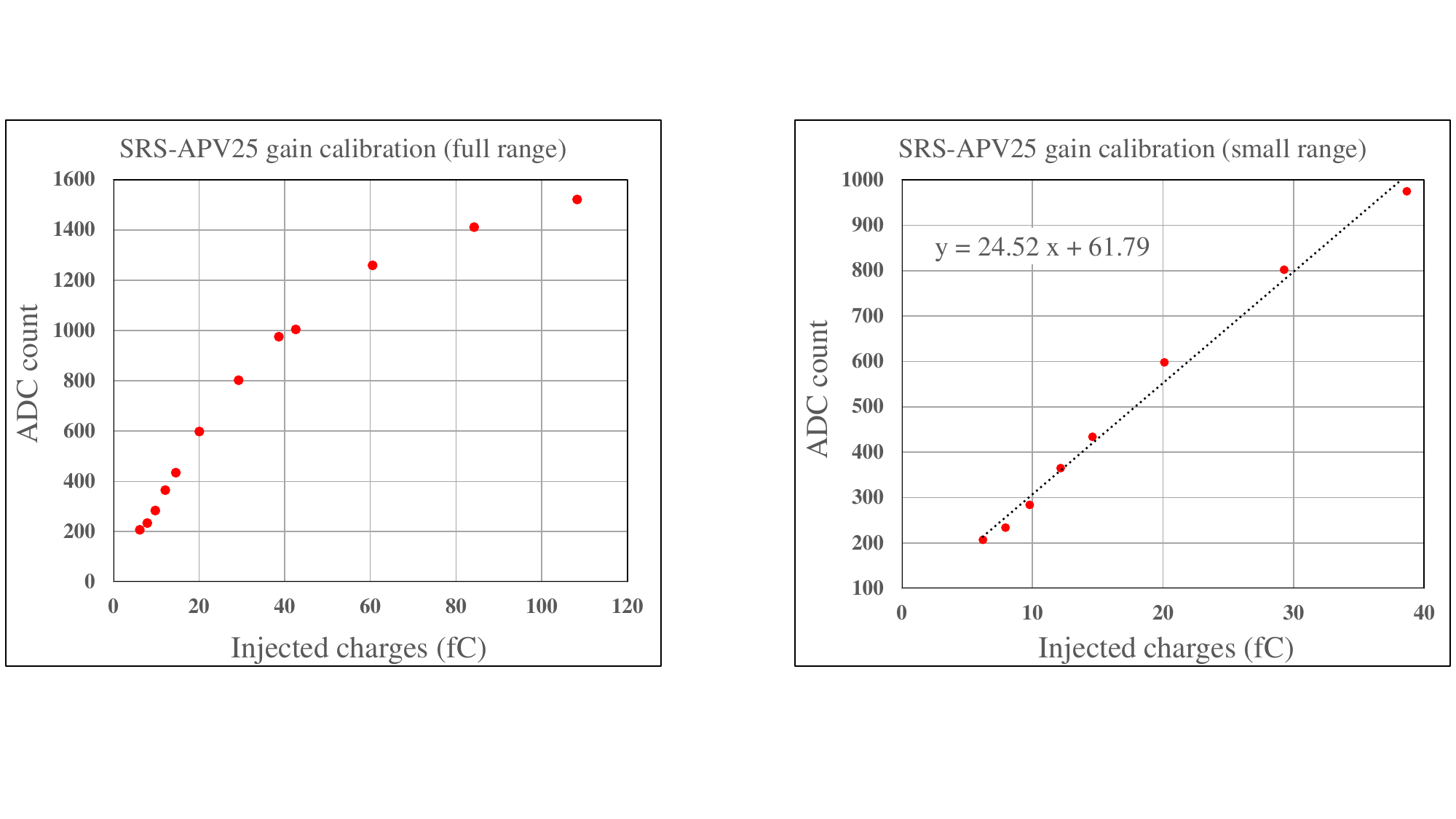}
\caption{\label{fig:apv_gain} SRS-APV25 gain calibration plots - ADC count vs. injected charges (fC) in a single APV channel; \textit{(Left:)} full range of the gain curve (non linear response  above $\sim$40 fC); \textit{(Right:)} the range of interest for the signal of the prototypes under study shows good linearity up to 40 fC.}
\end{figure}
The DATE package, developed by the ALICE collaboration at CERN~\cite{aliceSoft} and the amoreSRS~\cite{amoresrs} developed for the APV25-SRS readout  were used for the DAQ and data monitoring software respectively. The amoreSRS package includes scripts for raw APV25 data decoder, common mode correction, pedestal subtraction and zero suppression for online real-time monitoring as well as full offline analysis of the test beam data.\\
The gain calibration in ADC counts of the APV25-SRS signal amplitude is shown in the left plot of Fig.~\ref{fig:apv_gain} as a function of the injected charge on a single strip. The plot on the right shows a zoomed-in of the limited range of interest ($\le$40 fC, corresponding to $\sim$250,000 electron) for MPGDs where the APV25 shows a linear response to the injected charge. The range of linear response corresponds to an APV25 signal amplitude of 1000 ADC counts. A more detailed description of the gain calibration measurement is provided   in~\cite{apvgain}.  The gain of a typical MPGD detector is approximately 1 $\times \,10^4$, which corresponds to a total charge of $\sim$290,000 electrons in a 3~mm Ar:CO$_2$ (70:30) gas volume collected by the APV25 FE channels. If we assume equal sharing between between X-strips and Y-strips, each strip plane will collect on average  $\sim$145,000 electrons. Assuming that the average strip multiplicity per cluster is $\sim$4 for each readout plane and that central strip of the cluster picks up  $\sim$40\% of the total charge, the average charge of the central strip is $\sim$9 fC.  
%
%
\subsection{Strip pedestal noises}
\label{subsec:pedestals}
Dedicated pedestal runs were regularly taken between physics data runs during which the voltage applied to all detectors were lowered to prevent the development of avalanche signal in the detectors. The pedestal mean and rms values defining the pedestal offset and noise for each channel of APV25-SRS FE cards were then computed and stored in a ROOT~\cite{BRUN199781} file to be uploaded during the analysis to perform common mode correction, pedestal subtraction and zero suppression.
\begin{figure}[!ht]
\centering
\includegraphics[width=0.95\columnwidth,trim={0pt 0mm 0pt 0mm},clip]{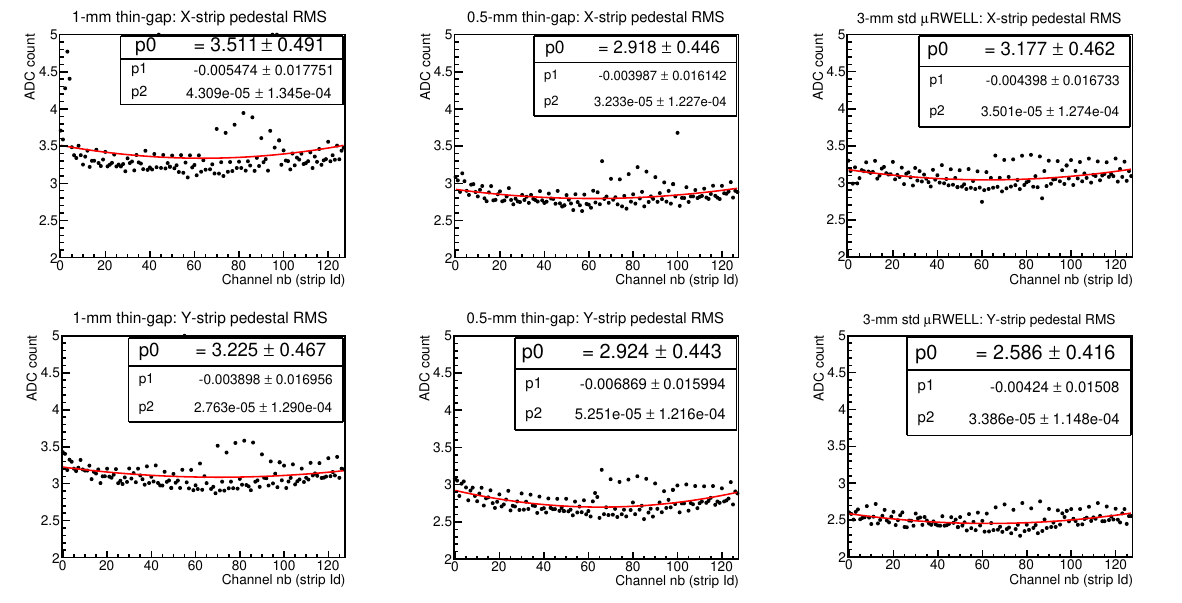}
\caption{\label{fig:pedestals} Plots of the pedestal noise (rms of the ADC distributions) for each the channel of APV25 FE cards connected to X-strips \textit{(top)} and Y-strips \textit{(bottom)} for  thin-gap proto-I \textit{(left)}; thin-gap proto-II \textit{(middle)}; and for reference $\muup$RWELL \textit{(right)}.} 
\end{figure}
The plots  in Fig.~\ref{fig:pedestals} show the pedestal noise in ADC counts expressed as the rms value of the pedestal fluctuation distributions for each channel of X-strips (top plots) and Y-strips (bottom plots) for thin-gap proto-I (left), thin-gap proto-II (center) and for the reference $\muup$RWELL (right). The average pedestal rms values  are $\sim$3.5 and $\sim$3.2 ADC counts in X-strips and Y-strips respectively for proto-I, $\sim$2.9 ADC counts in both X-strips and Y-strips for proto-II and  are $\sim$3.2 and $\sim$2.6 ADC counts in X-strips and Y-strips respectively for the reference $\muup$RWELL. For each triggered event, a zero suppression algorithm is applied to reject all APV25 channels with an ADC counts below a threshold defined by 5$\, \times \, \sigmaup_{ped} \,$ with $\sigmaup_{ped}$ being the pedestal noise (rms) of individual APV25 channel previously stored in the pedestal file. In this setup, the average threshold ranges from a minimum of $\sim$13~ADC counts for Y-strips of the reference $\muup$RWELL to a maximum of $\sim$17.5~ADC counts the X-strips of thin-gap proto-I. 
%
\subsection{FTBF 120 GeV beam profile}
\begin{figure}[!ht]
\centering
\includegraphics[width=0.995\columnwidth,trim={0pt 0mm 0pt 0mm},clip]{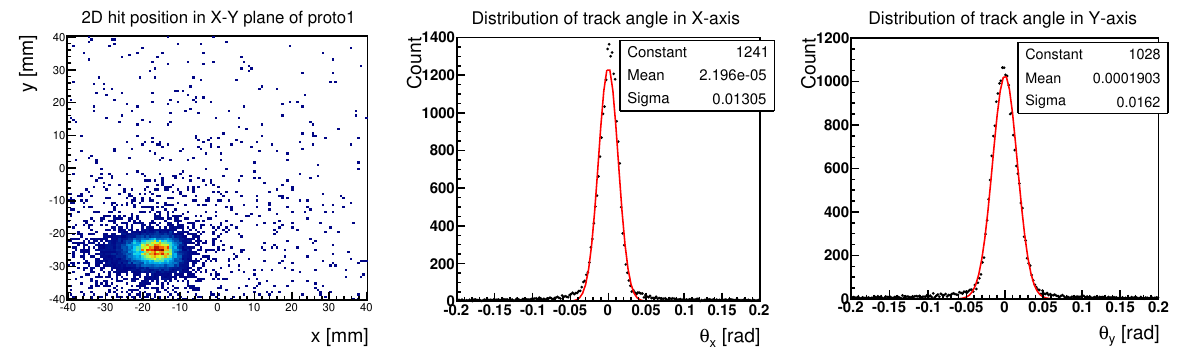}
\caption{\label{fig:beamprofile} \textit{(Left:)} 2D hit map from the reconstructed  coordinates x and y of the hits in thin-gap hybrid proto-I; angular distribution of the reconstructed tracks in x \textit{(center:)} and in y \textit{(right:)} from the hit coordinates in the four GEM trackers.}
\end{figure}

The plots  of Fig.~\ref{fig:beamprofile} show the 2D hit position map (left) of the proton beam impacting the detectors and the angular distributions of the reconstructed tracks in x-axis (center) and y-axis (right). A Gaussian function fit to the distributions shows the very narrow width of the angular distribution of the incoming proton beam, with a sigma equal to 0.013 radians (or  $\sim$0.75$^{\circ}$) in x-axis and 0.016 radians (or $\sim$0.9$^{\circ}$) in y-axis. This demonstrates that the proton beam is perpendicular to the X-Y plane of all detectors in the setup when the rotation stand of the prototypes is at home position (i.e. 0$^{\circ}$).

%% file: 04_hv_scan.tex
\section{HV scan studies}
\label{sec:hv_scan}
Four types of HV scan runs were conducted to evaluate the performance of the thin-gap GEM-$\muup$RWELL prototypes with regard to the average cluster charge distribution, the average strip multiplicity and the detector efficiency for both X-strips and Y-strips.  A brief description of the HV scan runs is as follows:
\begin{itemize}
  \item \textbf{GEM HV scan:} Variation of the voltage across the top and bottom electrodes of the GEM to study the impact of the pre-amplification on the performances.
  \item \textbf{$\muup$RWELL HV scan:} Variation of the voltage applied to the top Cu electrode of the $\muup$RWELL device with the DLC layer connected to the detector ground to study the impact of the second amplification on the performances.
  \item \textbf{Drift field HV scan:} Variation of the voltage across the drift region between the cathode and the top electrode of the GEM to study the impact of the electric field in the drift region  on the performances.
  \item \textbf{Transfer field HV scan:} Variation of the voltage in the transfer region between the bottom electrode of the GEM and the top electrode of the $\muup$RWELL to study the impact of the electric field in the transfer region on the performances.
\end{itemize}
A simple center of gravity (COG) algorithm is performed on a cluster of neighboring strips after the zero suppression steps described in~\ref{subsec:pedestals}, to reconstruct the particle position for both the X-strips and Y-strips. Single-strip clusters are included in the analysis script for all voltage scan studies. In the subsequent  sections, the data points plotted in red and blue, represent the results for the 1-mm  thin-gap GEM-$\muup$RWELL hybrid (proto-I) and for the 0.5-mm thin-gap GEM-$\muup$RWELL hybrid (proto-II) respectively. Solid square and open circle markers represent X-strips and Y-strips respectively while the solid and dashed lines are the fits to the data for X-strips and Y-strips respectively.
\subsection{\texorpdfstring{$\mu$}{}RWELL HV scan}
\begin{figure}[!ht]
\centering
\includegraphics[width=0.33\columnwidth,trim={0pt 0mm 0pt 0mm},clip]{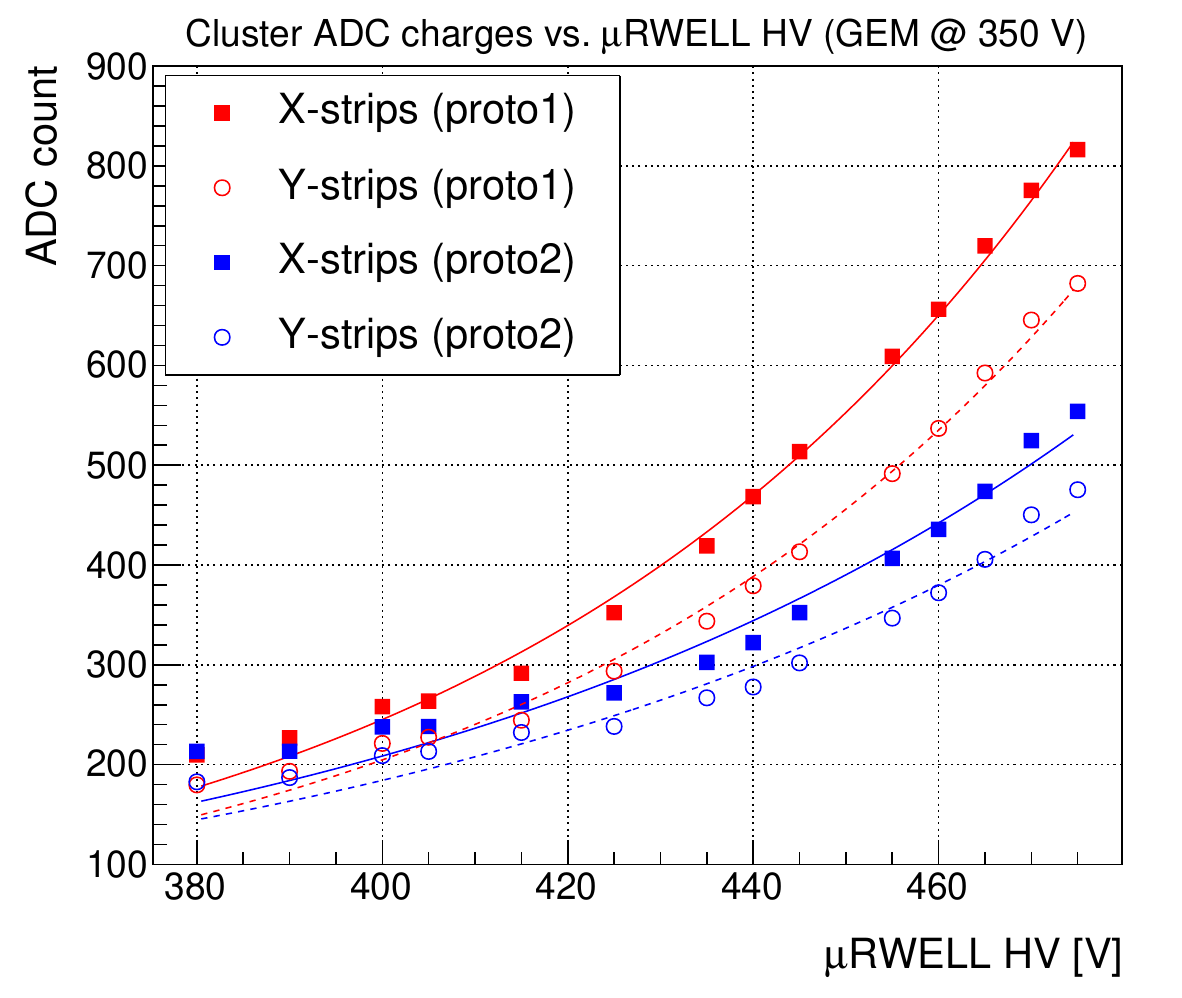}
\includegraphics[width=0.33\columnwidth,trim={0pt 0mm 0pt 0mm},clip]{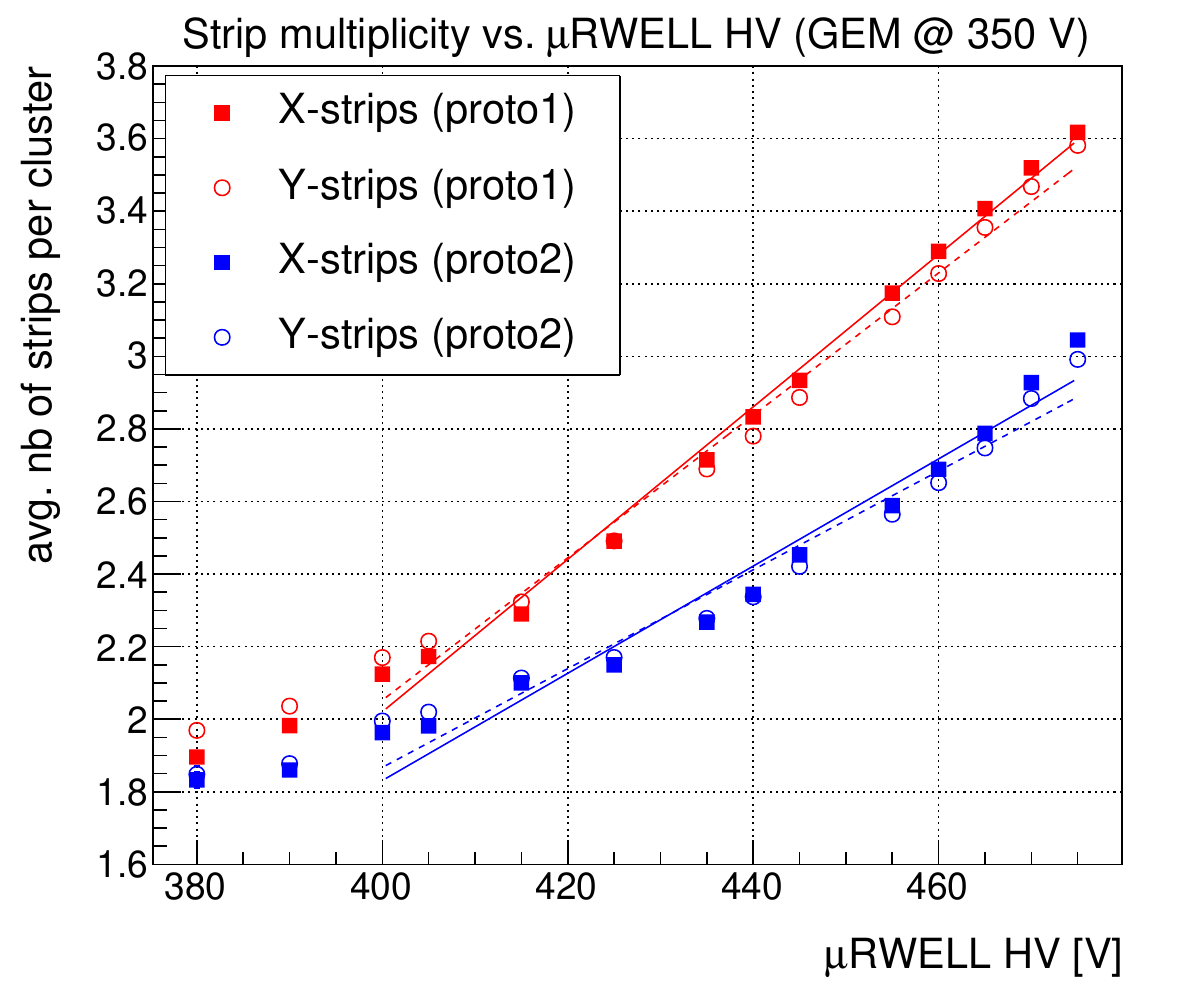}
\includegraphics[width=0.33\columnwidth,trim={0pt 0mm 0pt 0mm},clip]{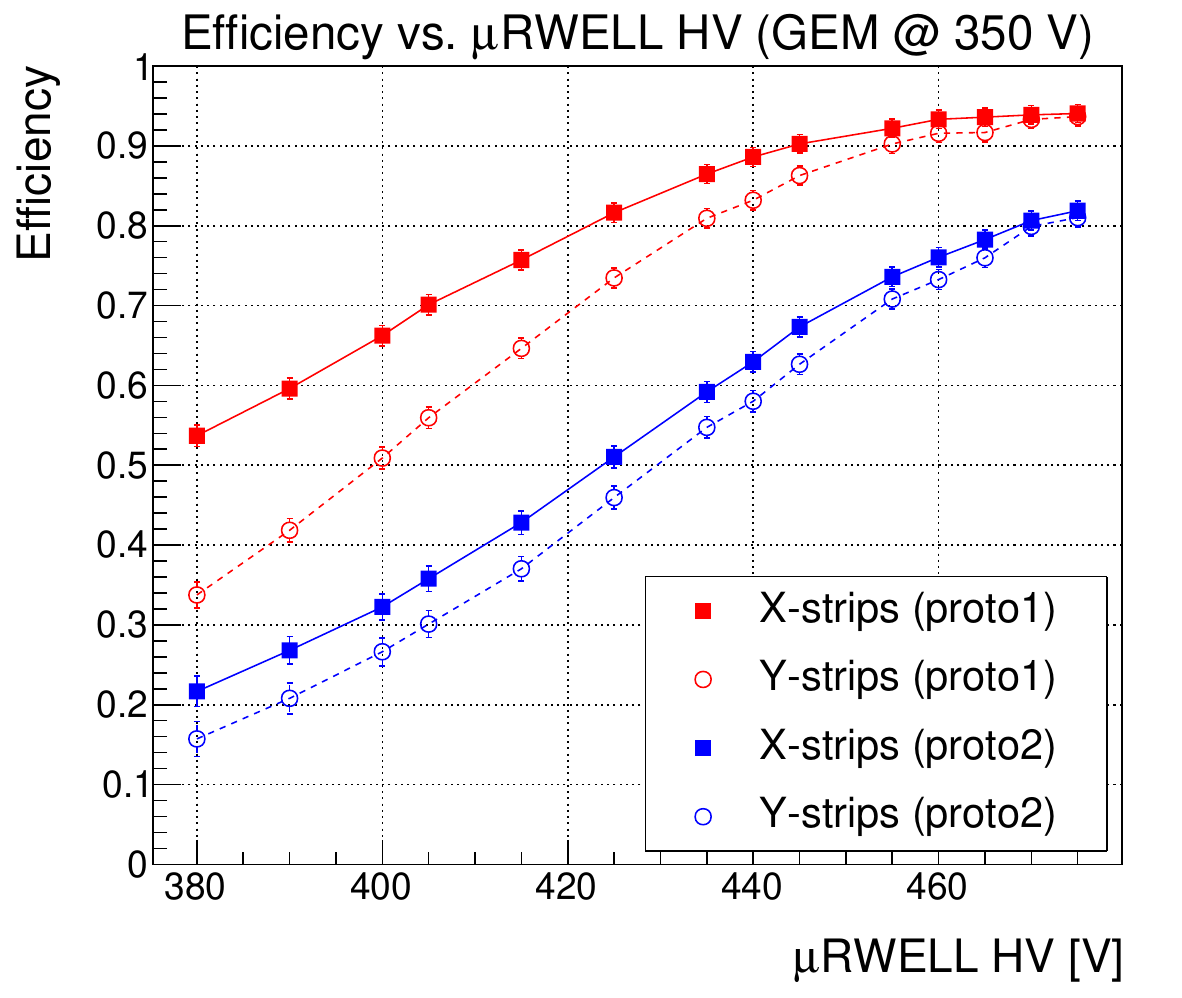}
\caption{\label{fig:jlab_urwellhv} HV scan of the bias voltage applied to $\muup$RWELL  foil -- impact of the second amplification on: \textit{(left:)} Average cluster charges in ADC counts; \textit{(center:)} average strip multiplicity; \textit{(right:)} detector efficiency per readout plane.}
\end{figure}

For the  $\muup$RWELL HV scan, the voltage applied to the top electrode of the $\muup$RWELL was varied in steps of 5~V or 10~V within the range of 380~V and 475~V. A constant voltage of 350~V was maintained across the GEM foil and  constant electric field (2~kV~/~cm) in the drift and transfer regions during the run. The plots on the left of Fig.~\ref{fig:jlab_urwellhv} show the average charge in the ADC channels in X-strips and Y-strips  for proto-I and proto-II plotted as a function of the $\muup$RWELL HV.   An exponential function fit to the data reflects the expected shape of gain curve of a gaseous detector resulting amplification as a function of the applied voltage. For both prototypes, the signal is  larger on  X-strips  than Y-strips and the difference is attributed to design of the X-Y strip readout layer which was not optimized for equal charge sharing as previously discussed in~\cite{capaSh_urwell2022}. Equal sharing could easily be achieved by properly adjusting the width of the strips of the top layer (X-strips) with respect to Y-strips of bottom layer but the charge sharing optimization is not the focus of this studies. As expected, at the same voltage setting, the signal for proto-I (in red) is significantly larger than for proto-II (in blue) because of the larger drift gap. \\ 
The average strip multiplicity, shown  in the center of Fig.~\ref{fig:jlab_urwellhv}   increases linearly with  $\muup$RWELL HV, reaching $\sim$3.6~strips for both X-strips and Y-strips for proto-I at a $\muup$RWELL HV setting of 470~V. For proto-II, even with a smaller signal amplitude and lateral size of the ionization charge cloud, we observe an average strip multiplicity of $\sim$3.0~strips for both X-strips and Y-strips. The efficiency plots in the X-strips and Y-strips for the two prototypes are shown on the graphs at right of  Fig.~\ref{fig:jlab_urwellhv}. The efficiency for proto-I  reaches a plateau at $\sim$94\% for both the X-strips and Y-strips at the voltage setting of 470~V.  The maximum efficiency expected from the simulation study of Fig.~\ref{fig:ionization}, is defined by the ratio of the number of events with zero ionization (content of first bin) to the total number of simulated events and is 1 - ($\sim$250 / 10,000) = $\sim$97.5\%. The 3.5\% discrepancy between simulation and experimental data can be attributed to the low amplitude signal events lost after zero suppression from the $\sim$10\% of the events that produced only one ionization in the gas volume. As expected, the efficiency is significantly lower at the same voltage setting for proto-II because of the smaller drift  gap of 0.5-mm.  The efficiency plateau starts around 81\% which once again is consistent with the simulation results of Fig.~\ref{fig:ionization} predicting a 15\% probability of zero ionization in a  0.5~mm Ar:CO$_2$ / 80:20. For proto-II, at the highest HV setting of 475~V, the full efficiency plateau has not been reached because of the limited beam test time.
\subsection{GEM HV scan}
\begin{figure}[!ht]
\centering
\includegraphics[width=0.33\columnwidth,trim={0pt 0mm 0pt 0mm},clip]{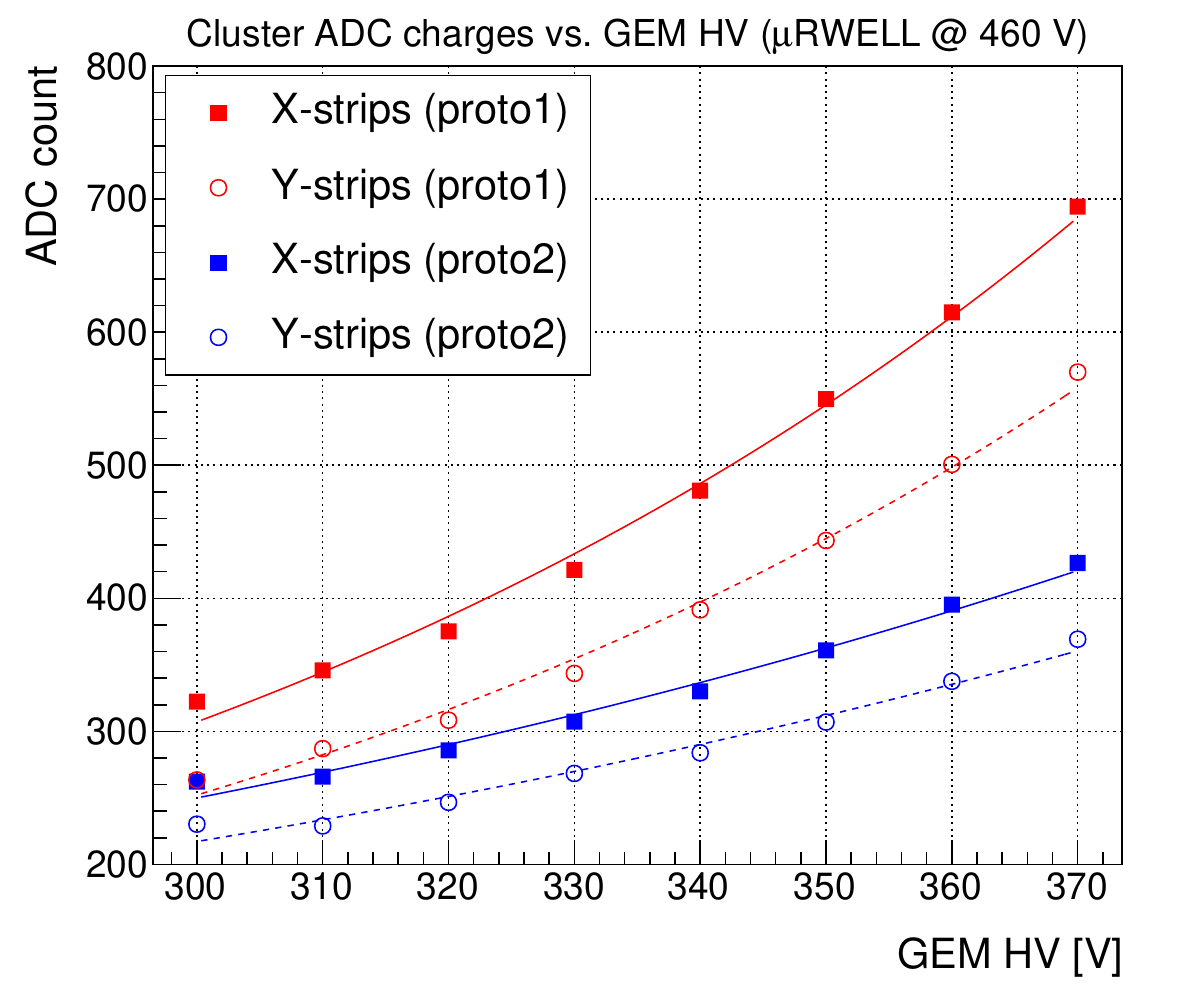}
\includegraphics[width=0.33\columnwidth,trim={0pt 0mm 0pt 0mm},clip]{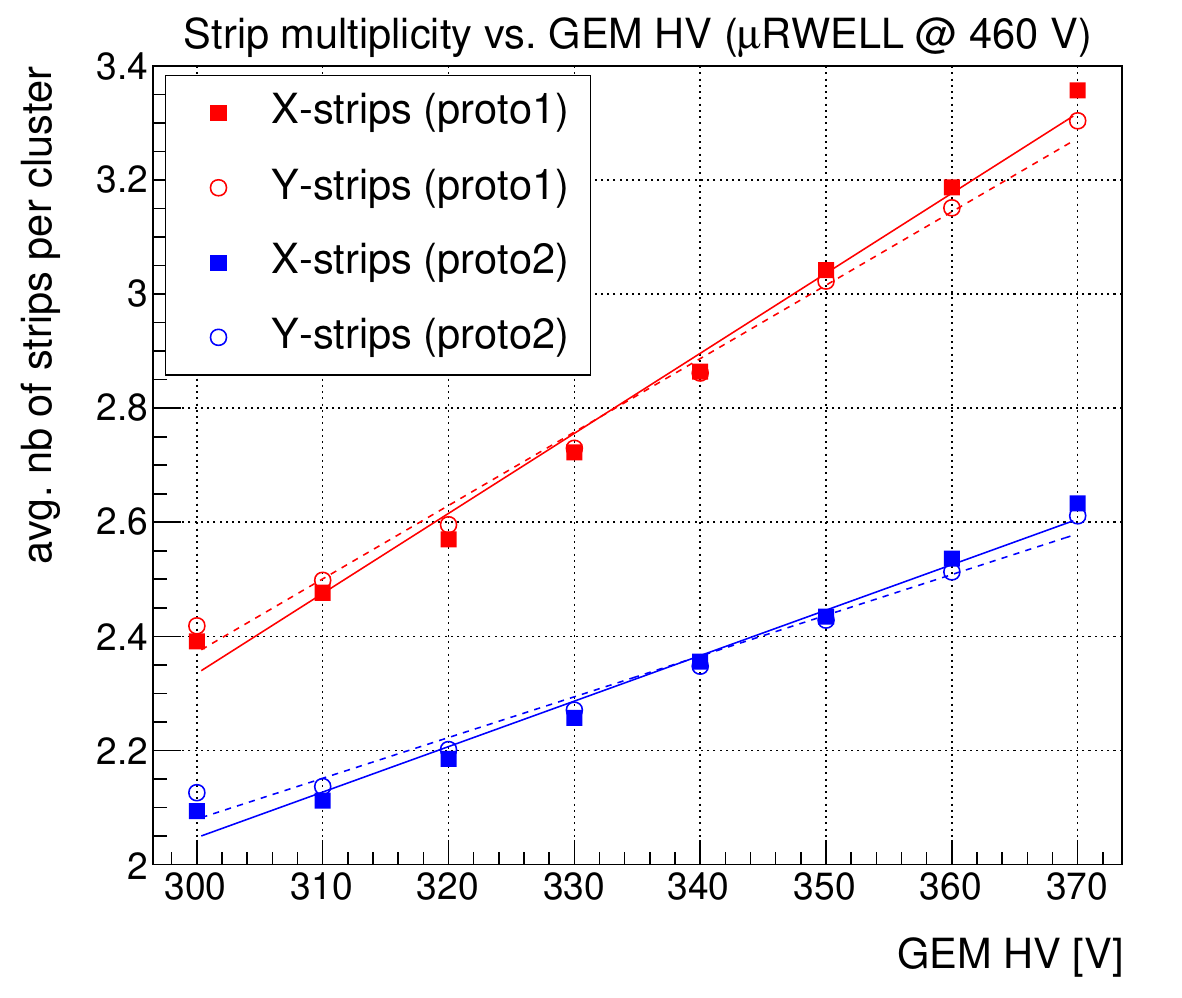}
\includegraphics[width=0.33\columnwidth,trim={0pt 0mm 0pt 0mm},clip]{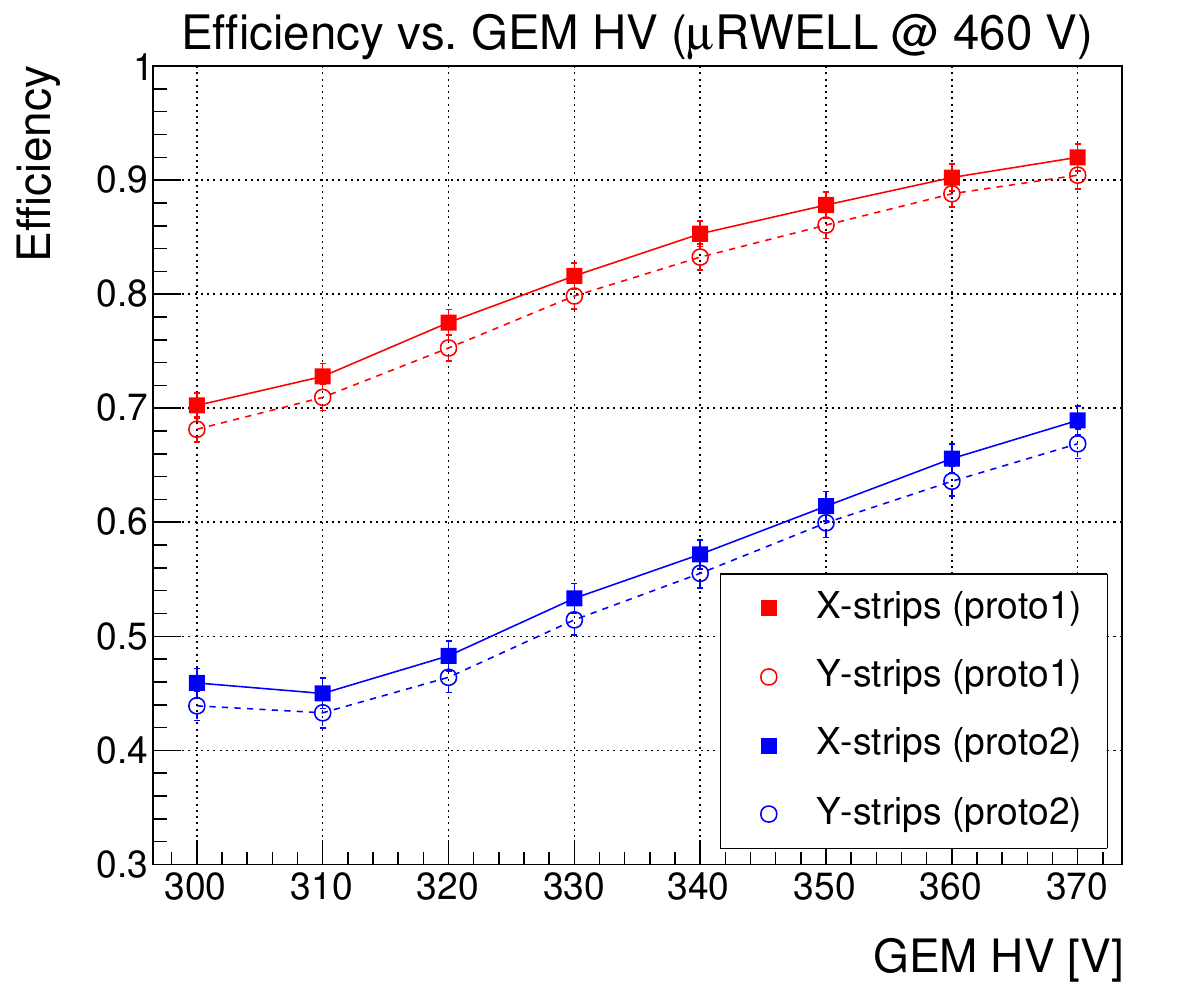}
\caption{\label{fig:jlab_gemhv} HV scan of the voltage across the GEM foil -- Study of the pre-amplification on: \textit{(left:)} Average cluster charges in ADC counts; \textit{(center:)} average strip multiplicity; \textit{(right:)} detector efficiency per readout plane.}
\end{figure}

For the GEM HV scans, the voltage, ranging from 300~V to 370~V, was applied across the top and bottom electrode of the GEM foil of the prototypes by steps of 10~V. The $\muup$RWELL bias voltage was maintained at 460~V and the electric field in both the drift and transfer regions of the chambers at  2~kV~/~cm. Here again, the ADC charges in X-strips and Y-strips increase exponentially with the GEM voltage as seen on the left of Fig.~\ref{fig:jlab_gemhv}. 
The amplitude of the signal is about 40\% smaller for the 0.5-mm gap prototype than the 1-mm gap at the larger voltage setting 370~V as shown in the plots. The strip multiplicity plots at the center of Fig.~\ref{fig:jlab_gemhv}, also exhibits a linear increase with the  GEM HV and reaches $\sim$3.3 on average in both X-strips and Y-strips at 370~V for proto-I and $\sim$2.6 on average for proto-II. As seen on the right panel of figure~\ref{fig:jlab_gemhv}, the efficiency of proto-I 
steadily increases from $\sim$70\% and $\sim$68\%  to $\sim$92\% and $\sim$90\% for  X-strips and Y-strips respectively. The maximum efficiency achieved for proto-II is  $\sim$70\% for X-strips and $\sim$68\% for Y-strips. 
\subsection{Transfer field scan}
\begin{figure}[!ht]
\centering
\includegraphics[width=0.33\columnwidth,trim={0pt 0mm 0pt 0mm},clip]{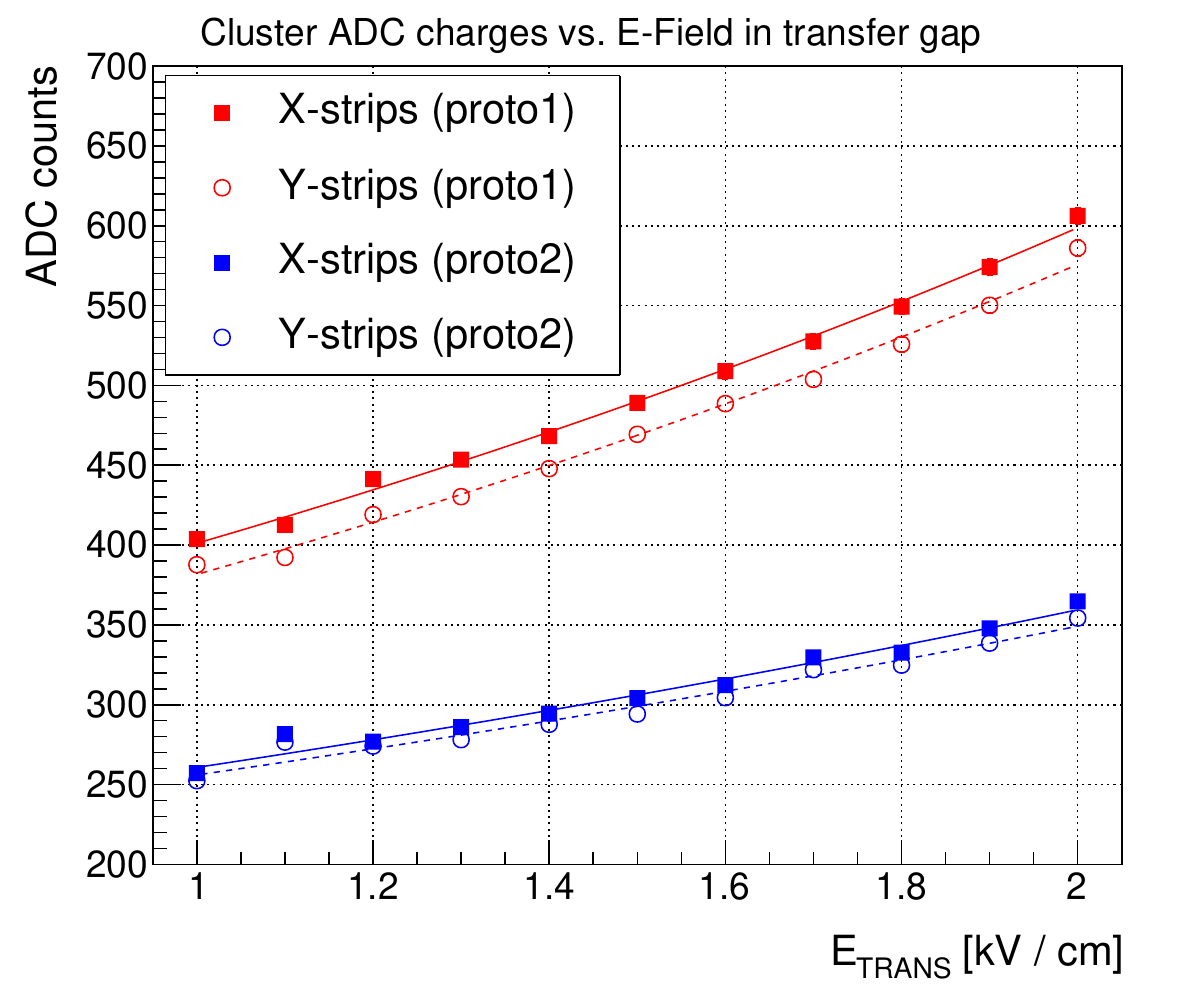}
\includegraphics[width=0.33\columnwidth,trim={0pt 0mm 0pt 0mm},clip]{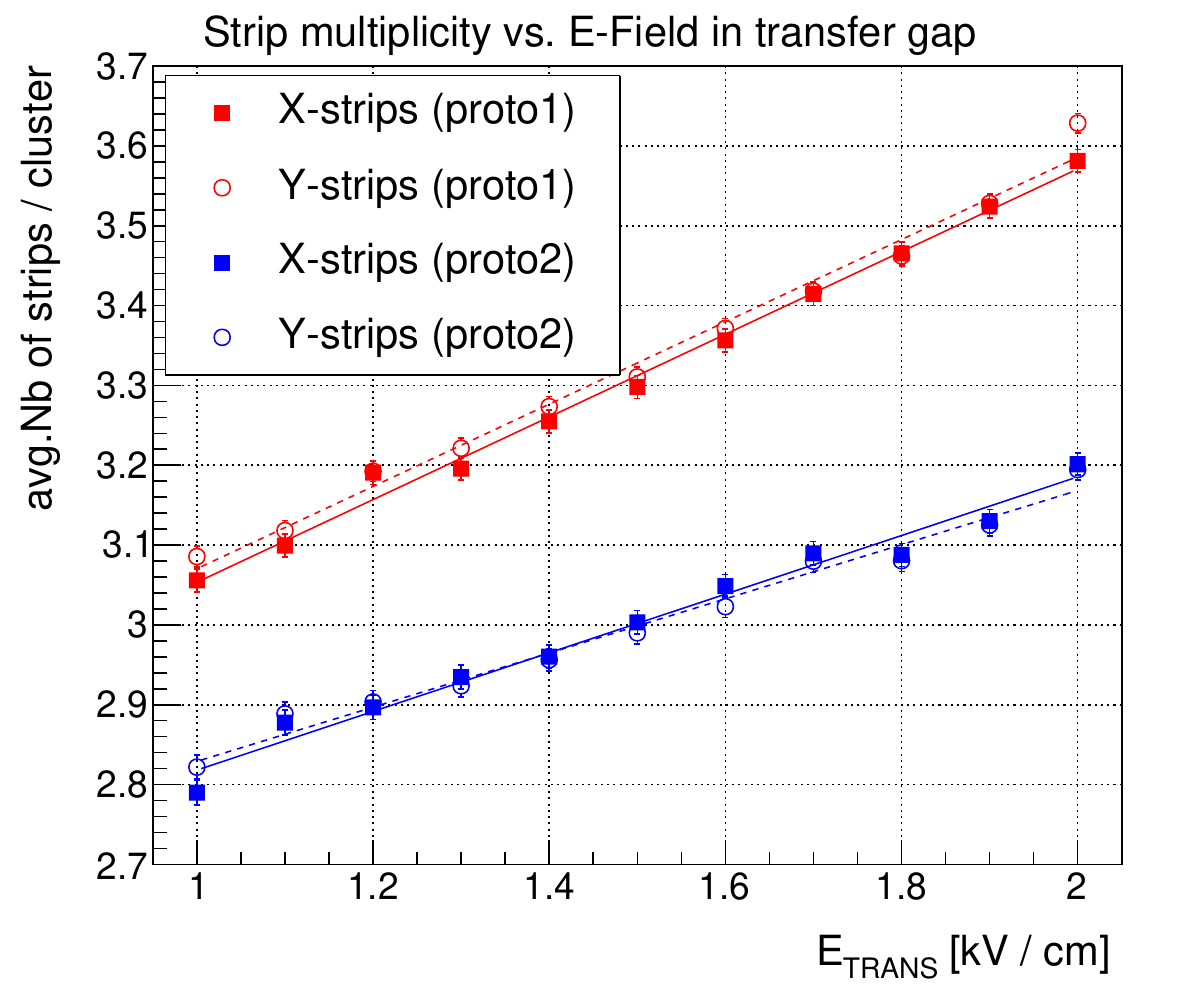}
\includegraphics[width=0.33\columnwidth,trim={0pt 0mm 0pt 0mm},clip]{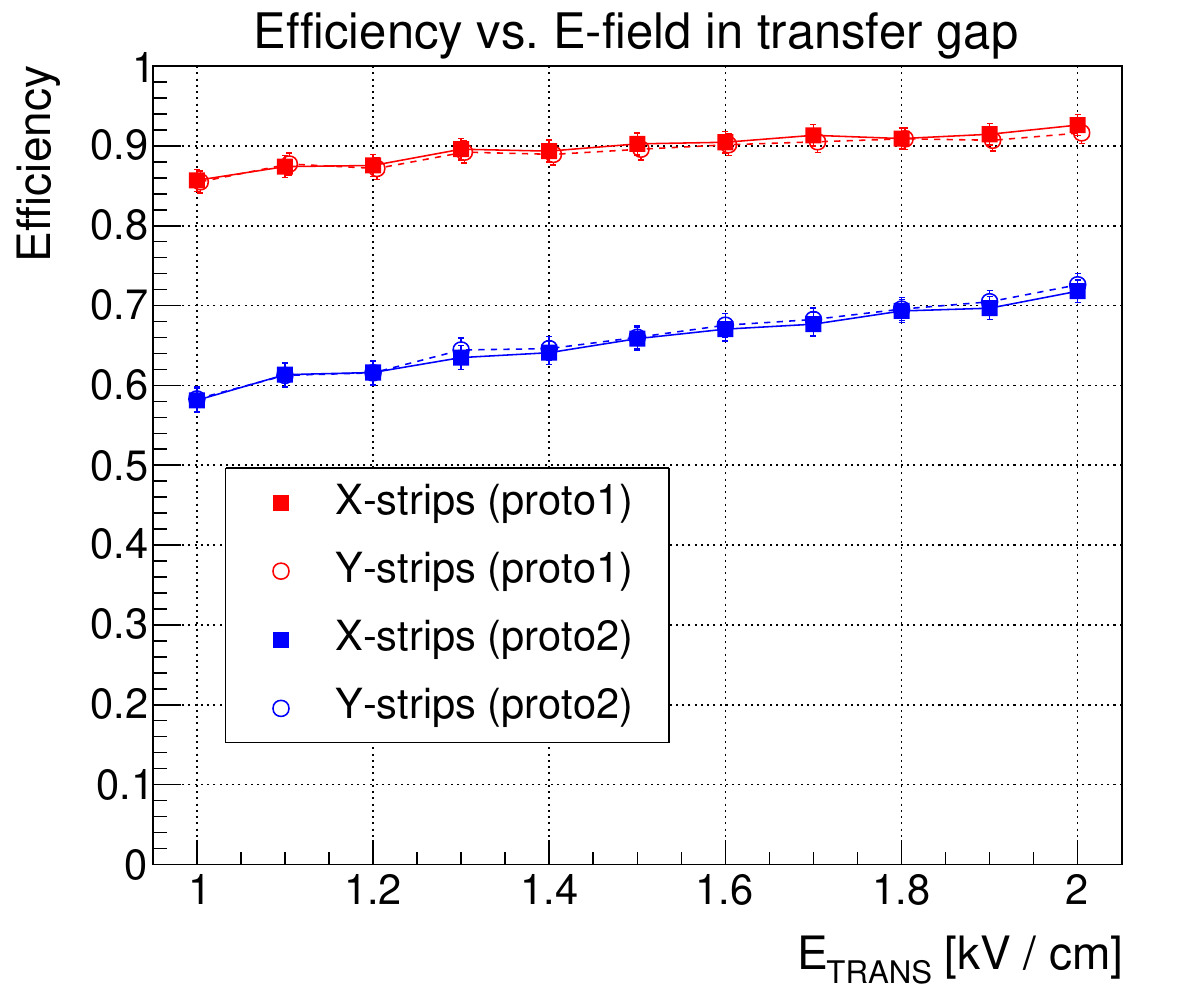}
\caption{\label{fig:jlab_indF} HV scan in the transfer region between the bottom of the GEM foil and the $\muup$RWELL PCB layer -- Impact of the transfer field ($\rm E_{TRANS}$) on: \textit{(left:)} Average cluster charges in ADC counts; \textit{(center:)} average strip multiplicity; \textit{(right:)} detector efficiency per readout plane.}
\end{figure}

The electric field in the transfer region has a significant impact on the performance of the prototypes as seen in the plots of  Fig.~\ref{fig:jlab_indF}. For this run, the voltages on the  $\muup$RWELL and across the GEM foil were kept constant at 460~V and 350~V respectively with the electric field in the drift region at 2~kV~/~cm while varying the voltage in the transfer region of the prototypes between the bottom electrode of the GEM foil and the top electrode of  $\muup$RWELL from 100~V to 200~V in steps of 10~V corresponding to  the electric field variation of 1~kV~/~cm to 2~kV~/~cm.
 The left panel of Fig.~\ref{fig:jlab_indF} shows the dependence of the signal amplitude on the transfer field. The amplitude of the signal increased linearly from $\sim$400~ADCs on average for both X-strips and Y-strips to $\sim$600~ADCs when the electric field increases from 1~kV~/~cm to of 2~kV~/~cm in the transfer gap. Similar increase is observed for proto-II (blue curves). This represents a 50\% increase of the signal amplitude when the strength of the electric field in the transfer gap is doubled.  Strip multiplicity also increases steadily by roughly 10\% to 15\% with the field in the transfer gap for both prototypes as seen in the plots in the center of  Fig.~\ref{fig:jlab_indF}, along with the steady increase of efficiency for proto-I from $\sim$86\% to $\sim$92\% for both X-strips and Y-strips and from     from $\sim$58\% to $\sim$72\%  for proto-II as seen in the right of the figure. 
\subsection{Drift field scan}
A scan of the electric field in the drift region was also performed by varying the voltage applied to the cathode with respect to the voltage at the top electrode of the GEM foil.
\begin{figure}[!ht]
\centering
\includegraphics[width=0.33\columnwidth,trim={0pt 0mm 0pt 0mm},clip]{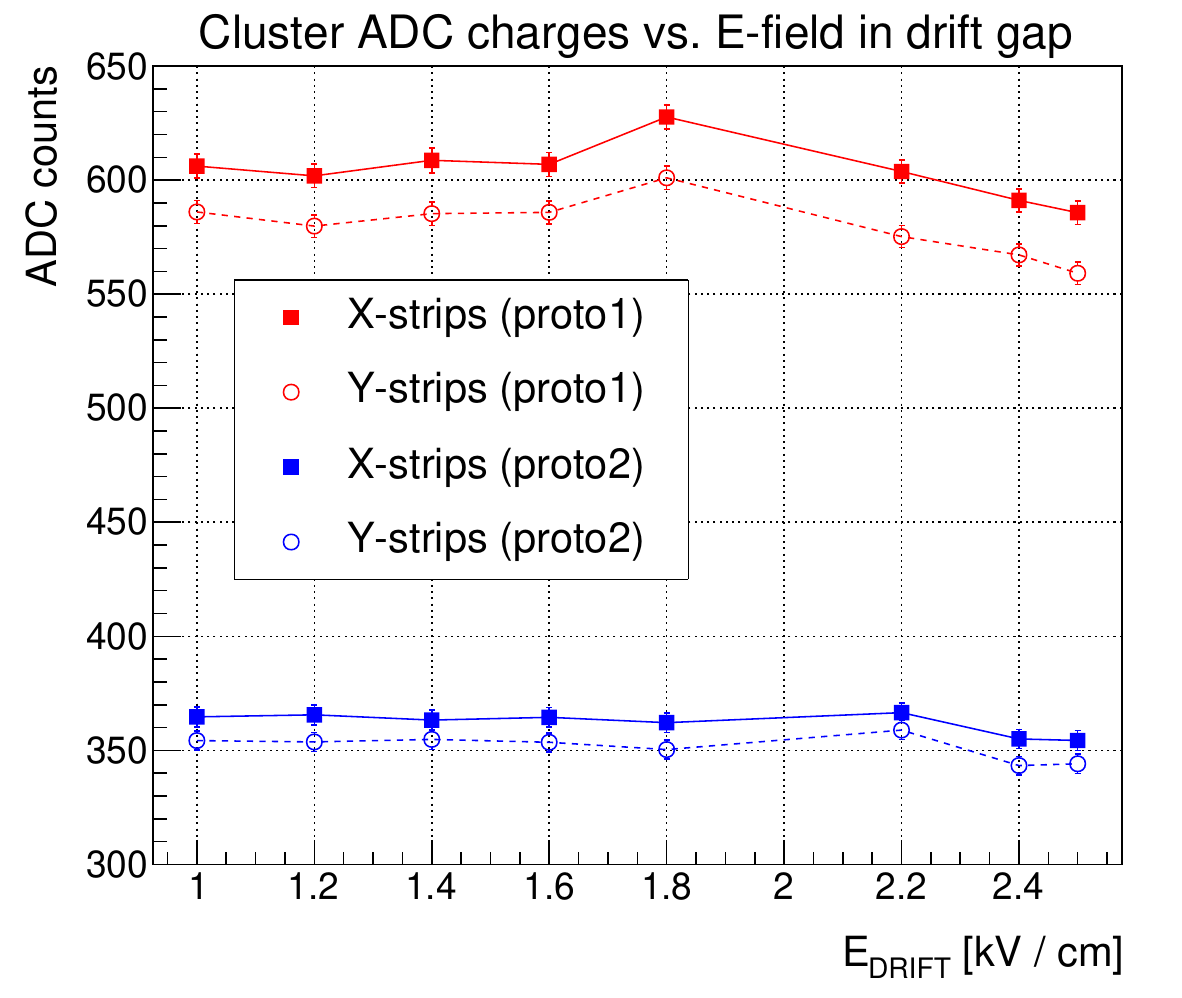}
\includegraphics[width=0.33\columnwidth,trim={0pt 0mm 0pt 0mm},clip]{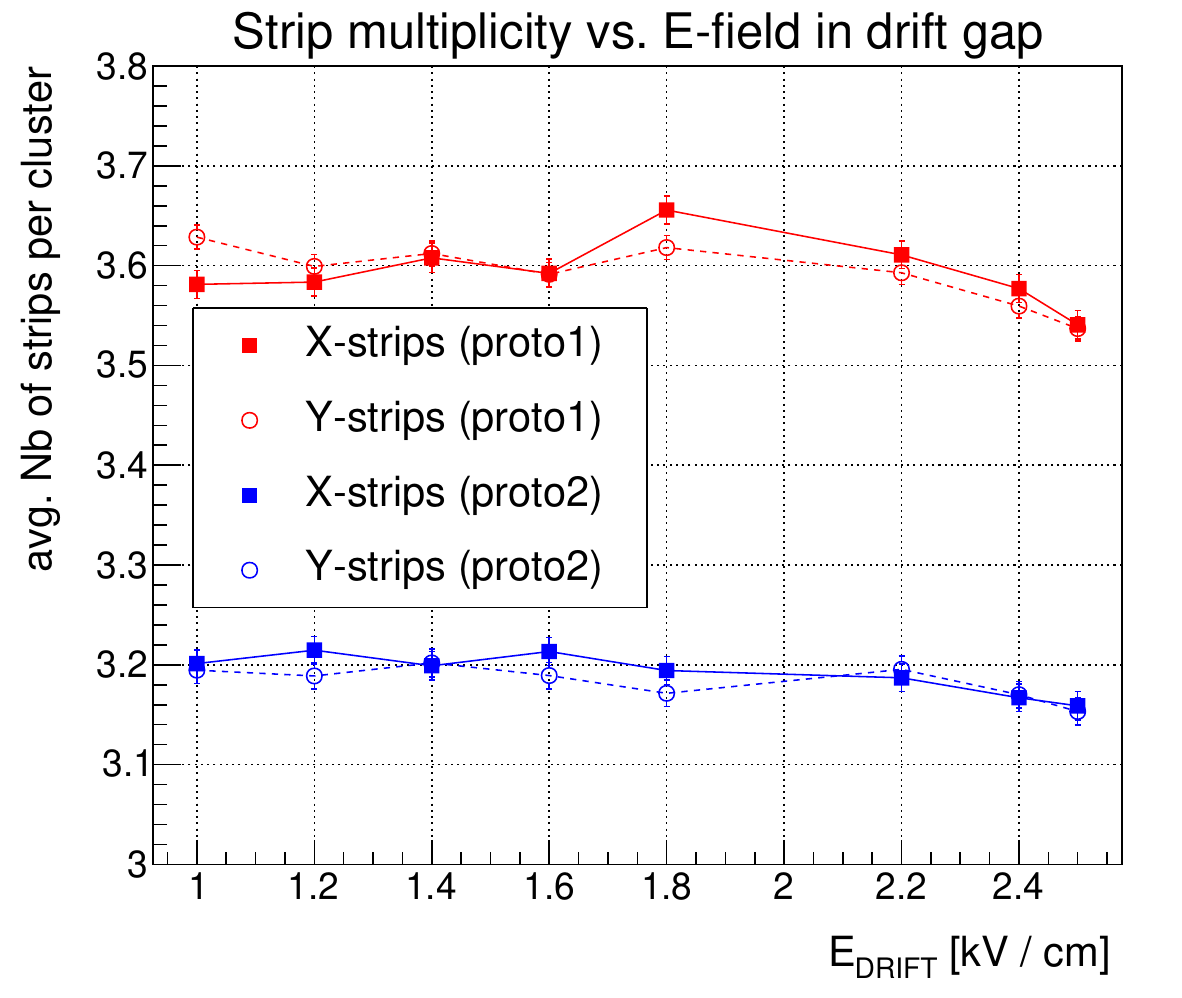}
\includegraphics[width=0.33\columnwidth,trim={0pt 0mm 0pt 0mm},clip]{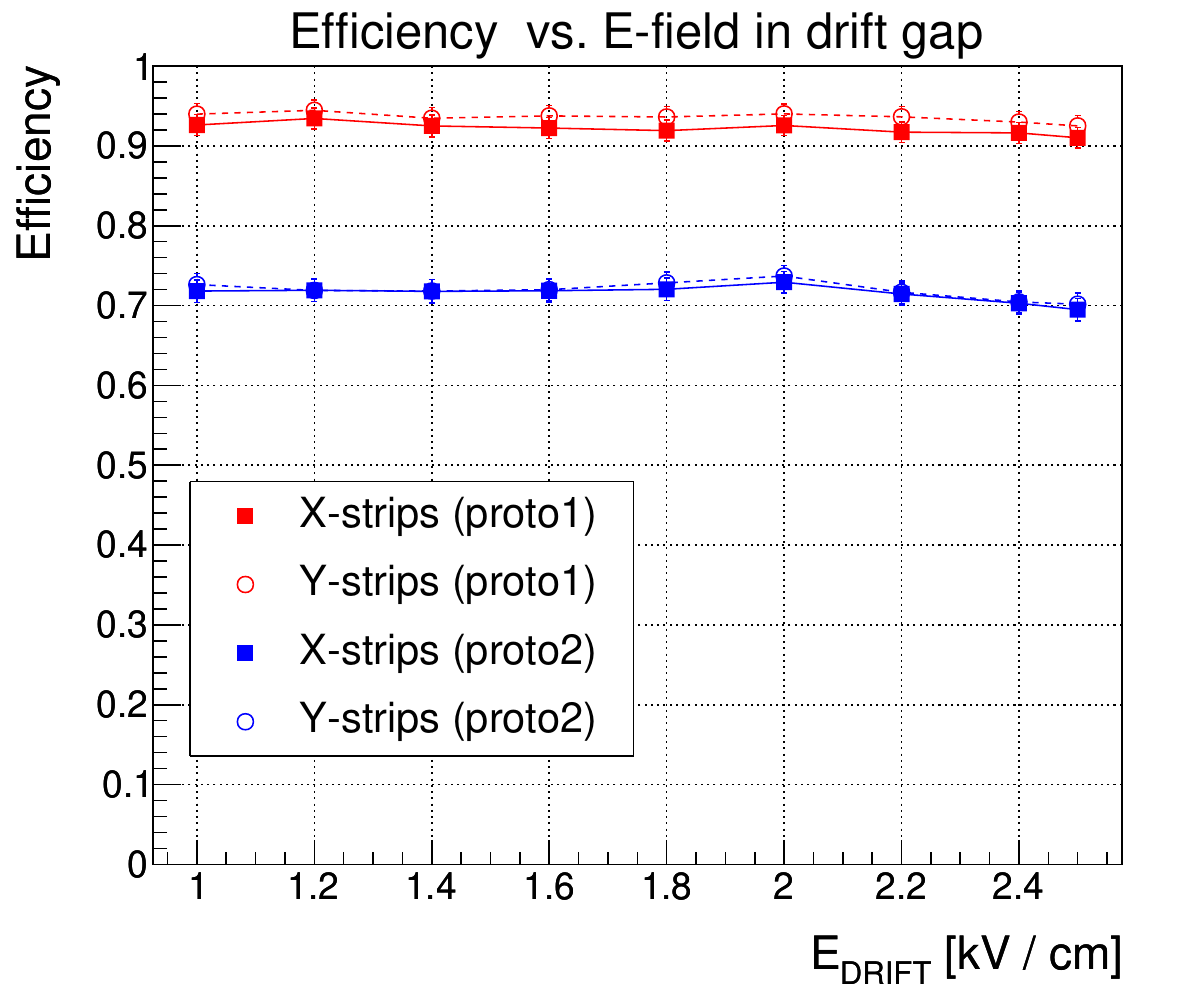}
\caption{\label{fig:jlab_driftF} HV scan in the drift region between the cathode foil and the top electrode of the GEM foil -- Impact of the drift field ($\rm E_{DRIFT}$) on: \textit{(left:)} Average cluster charges in ADC counts; \textit{(center:)} average strip multiplicity; \textit{(right:)} detector efficiency per readout plane.}
\end{figure}
The field was increased from 1~kV~/~cm in the drift region to 2.5~kV~/~cm in steps of 200~V~/~cm. The voltage on the  $\muup$RWELL and GEM foil were kept at 460~V and 350~V respectively and the electric field in the transfer region was set to 2~kV~/~cm. As shown in the plots of  Fig.~\ref{fig:jlab_driftF},  there was no  significant impact of the drift field in the performance of the detector in this electric field range of the study. The efficiency fluctuates around  92\% for proto-I for both X-strips and Y-strips and $\sim$72\%  for proto-II. 

%% file: 05_angle_scan.tex
\section{Angle scan studies}
\label{sec:angle_scan}
A dedicated angle scan run was performed to compare the performance of the two thin-gap proto-I and proto-II with a standard (3~mm drift gap)  $\muup$RWELL prototype used for reference. The three prototypes were mounted on the rotation stand described in section~\ref{subsec:rotstand} for the rotation of the detector X-Y plane along the vertical Y-axis in a range of 0$^{\circ}$ to 45$^{\circ}$ incrementing by 3$^{\circ}$ with respect to the proton beam direction along z axis. For this run, the  voltages on the $\muup$RWELL and across the GEM were 460~V and 350~V respectively and the electric field in the drift and transfer gap were set to 2~kV~/~cm for the two thin gap prototypes. The voltage applied to the reference $\muup$RWELL prototype was 575~V and the electric field in the drift gap was equal to 2~kV~/~cm. The signal amplitude, the average strip multiplicity, the detector efficiency and the spatial resolution were  measured as a function of track angle for the prototypes.  
\subsection{Signal amplitude, strip multiplicity  and efficiency vs. track angle}
\begin{figure}[!ht]
\centering
\includegraphics[width=0.33\columnwidth,trim={0pt 0mm 0pt 0mm},clip]{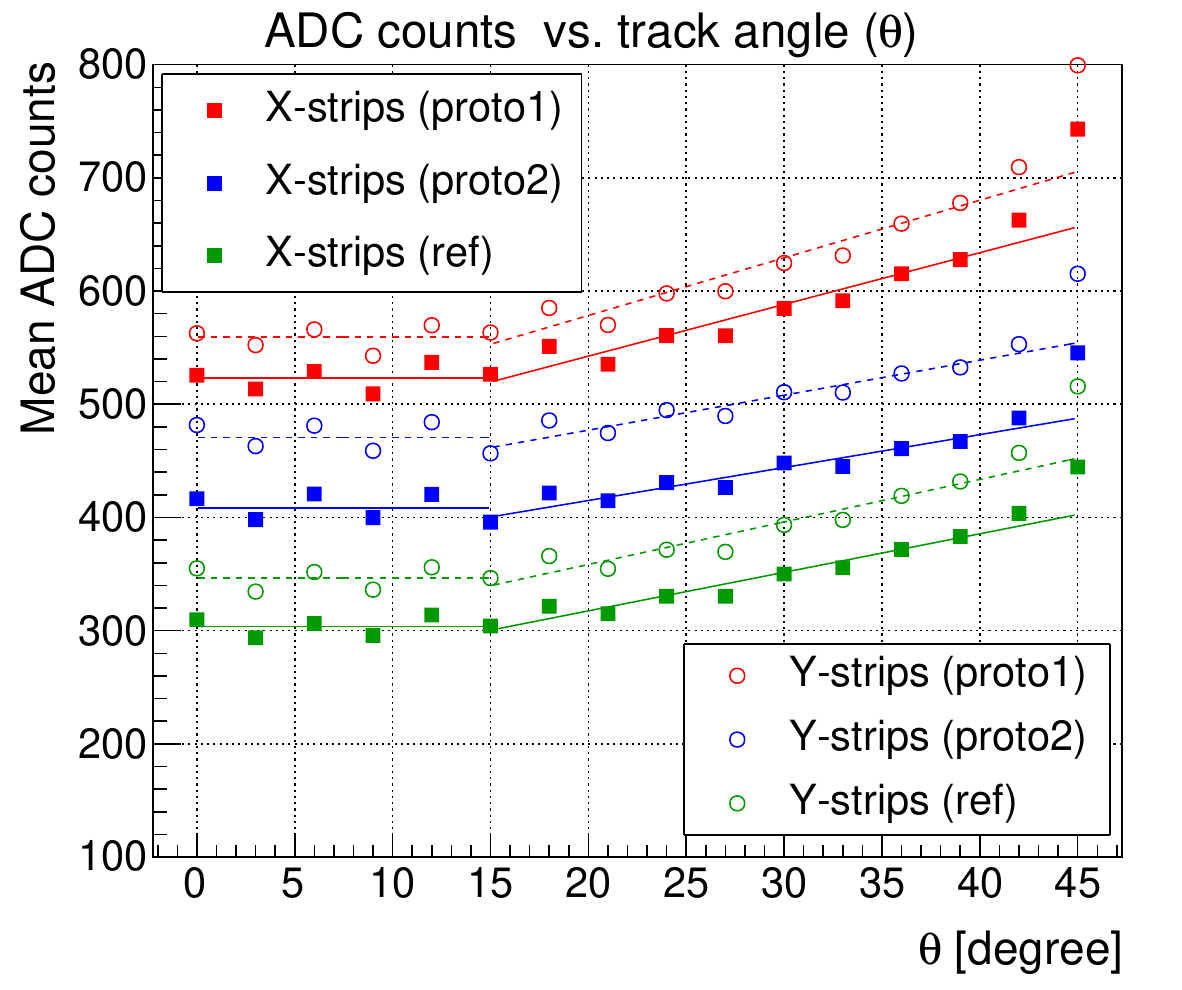}
\includegraphics[width=0.33\columnwidth,trim={0pt 0mm 0pt 0mm},clip]{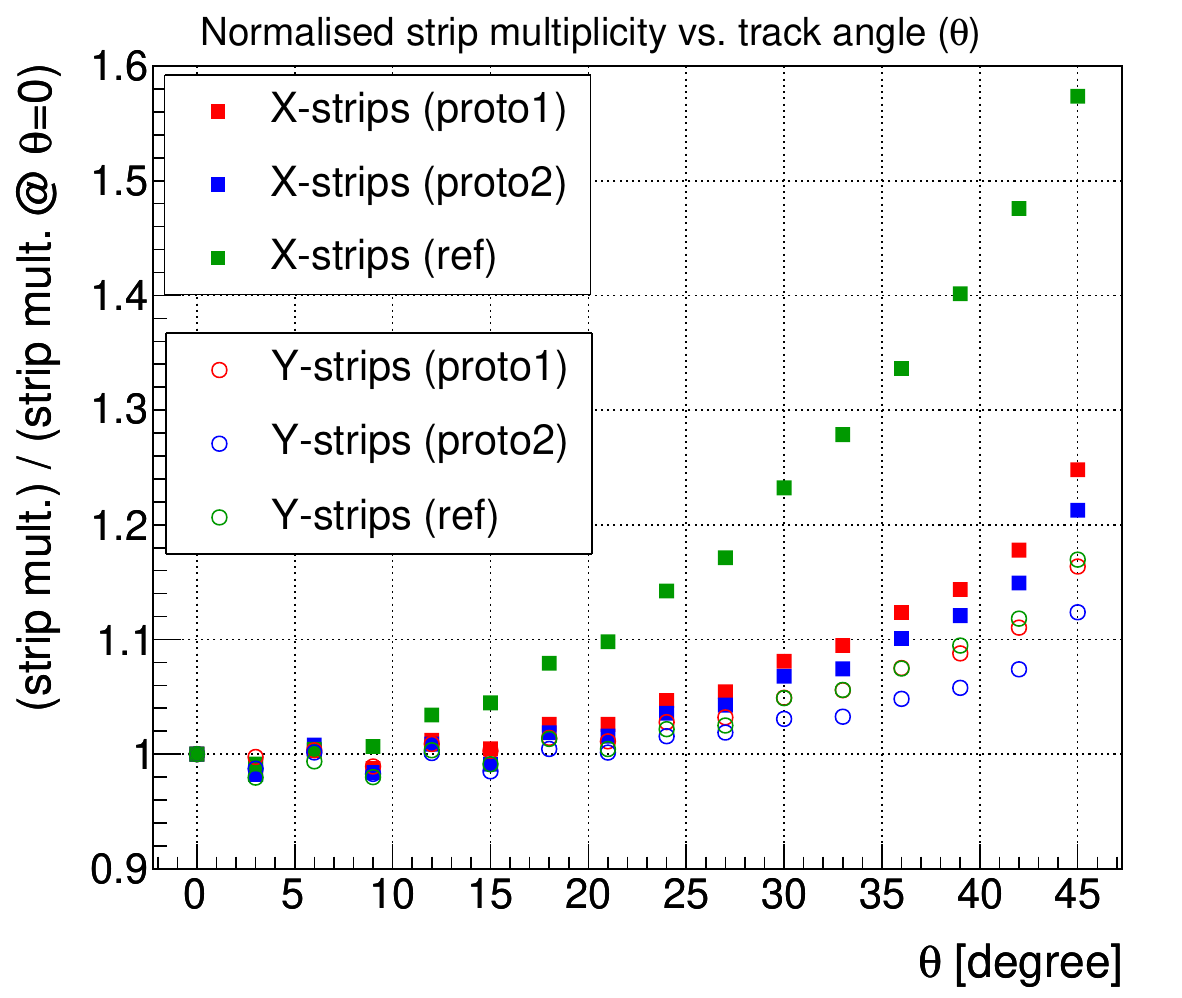}
\includegraphics[width=0.33\columnwidth,trim={0pt 0mm 0pt 0mm},clip]{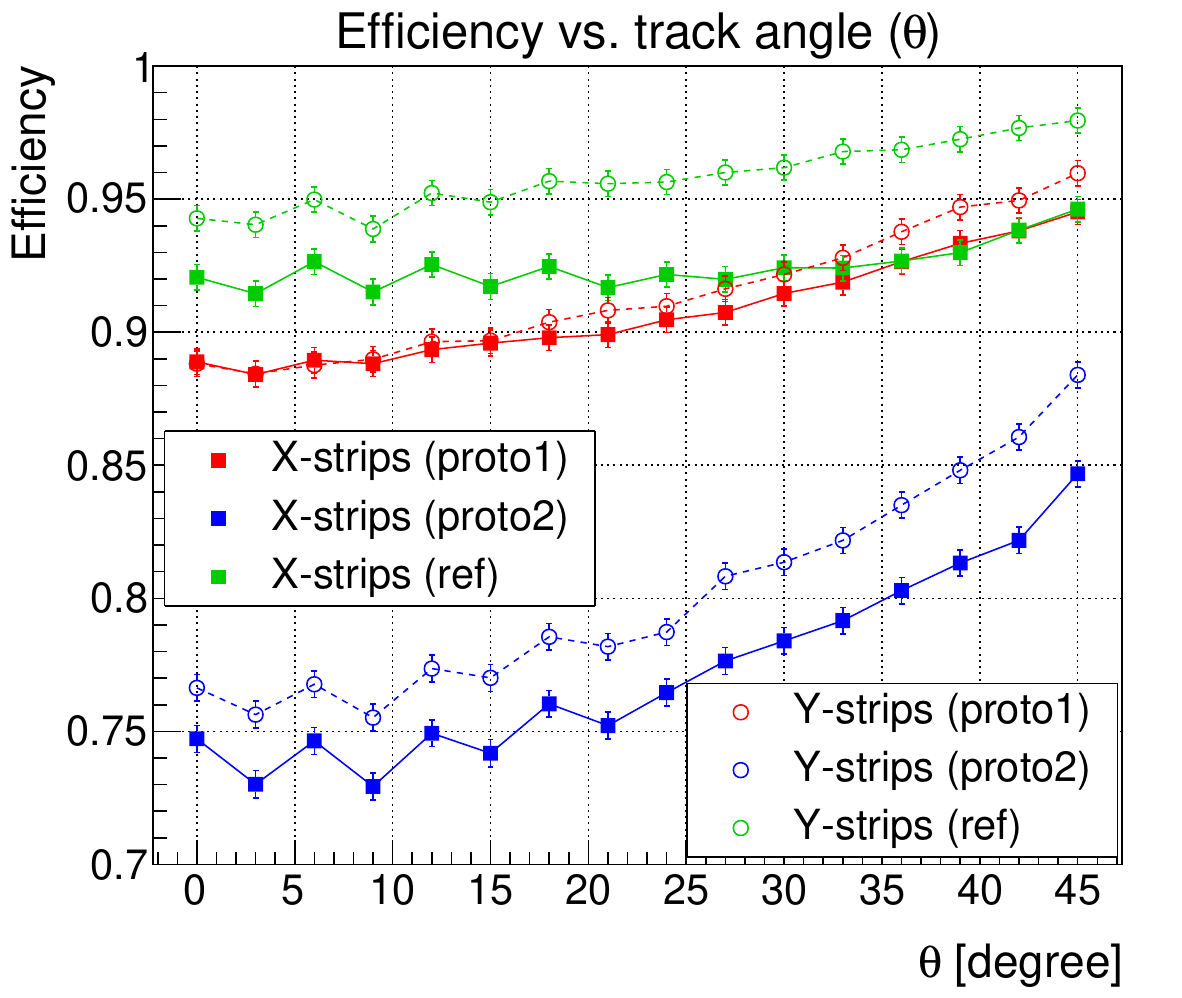}
\caption{\label{fig:jlab_eff_vs_ang} (\textit{left:}) Signal amplitude in ADC counts  vs. track angle; (\textit{center:}) Average strip multiplicity normalized to the strip multiplicity at  0$^{\circ}$ vs. track angle; (\textit{right:}) Efficiency vs. track angle;}
\end{figure}

The left panel of the Fig.~\ref{fig:jlab_eff_vs_ang} shows the average charge in ADC counts as a function of the angle of the incoming particles in X and Y for all three prototypes. The average ADC is nearly constant in the angle range of 0$^{\circ}$ to $\sim$15$^{\circ}$ but  increases linearly for impact angles $\geq$ 15$^{\circ}$,  up to 39$^{\circ}$.  
The charge increase is correctly correlated to the length of ionization trail of the impinging particles in the drift region. The graph in the middle panel  shows the average strip multiplicity plotted as a function of the track angle for X-strips and Y-strips and normalized to their value for perpendicular tracks at 0$^{\circ}$.\\
The normalized strip multiplicity plot provides a better idea of the impact of track angle on the strip multiplicity for the prototypes under study. In these plots, we can clearly observe a strong correlation between the drift gap and the relative increase of average strip multiplicity for X=strips. For the reference $\muup$RWELL with 3~mm drift gap, the relative strip multiplicity increases by $\sim$60\% (green square dots) when the X-Y plane is rotated from 0$^{\circ}$ to 45$^{\circ}$. The relative increase is significantly less for the thin gap detectors, reaching  $\sim$25\% for proto-I (red square dots) and $\sim$21\% for proto-II (blue square dots). The increase of strip multiplicity for X-strips is dominated by the longer ionization paths of the incoming particle in this axis, leading to more strips with hit above threshold and contributing to the degradation of the spatial resolution as shown in  section~\ref{subsec:jlab_trk_angle_scan}.  
For  Y-strips (open squares), the relative increase of the strip multiplicity reaches $\sim$18\% at 45$^{\circ}$ for proto-I and the reference $\muup$RWELL detector and  $\sim$14\% for proto-II. Unlike the situation for the X-strips, the particle tracks remain normal to the detector in Y-direction and the increase in Y-strips is mainly due to the larger signal amplitude from to the total ionization charge which increases proportionally with the track angle as shown in the plots of the left panel. 
The dependence of detector efficiency on the angle of incoming tracks is shown on the right panel plot of Fig.~\ref{fig:jlab_eff_vs_ang} for the two prototypes and the reference detector. The efficiency steadily increases with the angle in both readout planes of each detector. For proto I, the efficiency starts slightly below 90\% for tracks at $\thetaup\,=\,0^{\circ}$  and close to 95\% and 96\%  for $45^{\circ}$ tracks in both X-strips and Y-strips respectively, in Ar:CO$_2$ gas mixture. Thus the average detector efficiency for proto-I is  $\sim$93\% for the  track angle range of $0^{\circ}$ to $45^{\circ}$. For proto-II the efficiency increases from 73\% from 85\% for X-strips and 76\% from 87\%  for X-strips for tracks at $0^{\circ}$ and $45^{\circ}$ respectively. 
\subsection{Track fit residuals}
\label{sec:jlab_trk_residuals}%
\begin{figure}[!ht]
\centering
\includegraphics[width=0.995\columnwidth,angle=0, trim={0pt 0mm 0pt 0mm},clip]{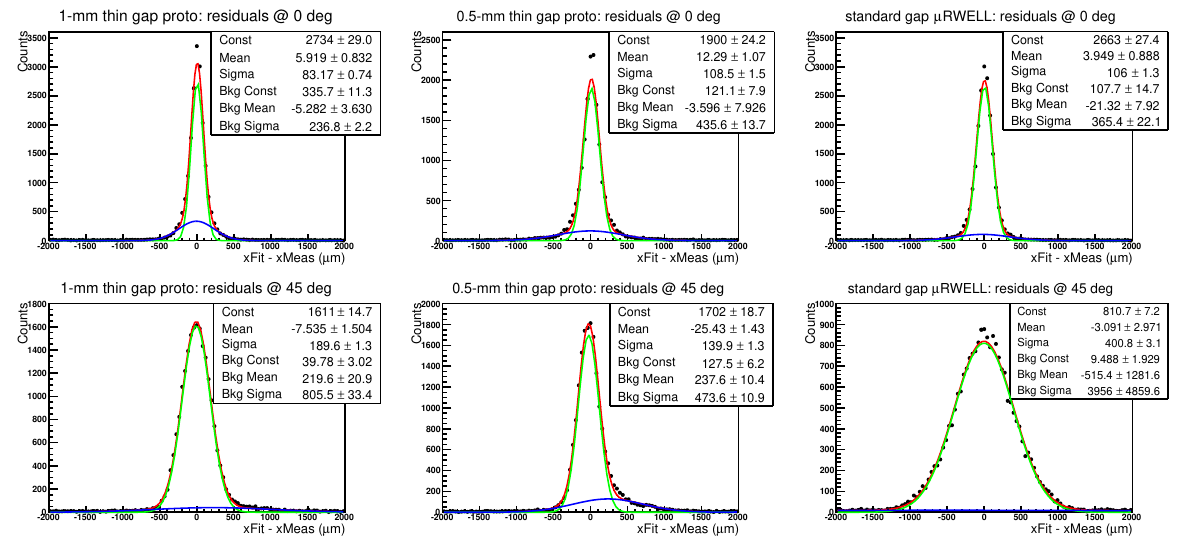}
\caption{\label{fig:jlab_res} Track fit residuals distribution in X-strips for proto-I (\textit{left}), proto-II (\textit{center}) and for reference  (\textit{right}) for  tracks perpendicular to the detector planes $\thetaup\,=\,0^{\circ}$ (\textit{top}) and for tracks at an angle  $\thetaup\,=\,45^{\circ}$ (\textit{bottom}).}
\end{figure}

The plots in Fig.~\ref{fig:jlab_res} show the distribution of the residuals from the fitted tracks in X-strip for proto-I (left),  proto-II (middle) and for reference $\muup$RWELL (right). Top plots show the residuals for perpendicular tracks at the angle $\thetaup\,=\,0^{\circ}$ and bottom plots, the residuals at the angle $\thetaup\,=\,45^{\circ}$. The data are fitted by a double Gaussian function (red curves) in the plots. The width of the narrow Gaussian function (green) is used for the residual width and by extension for the spatial resolution of the detector.
The width of the residuals for X-strips is 83.17~$\muup$m, 108.5~$\muup$m and  106~$\muup$m for proto-I, proto-II and the reference $\muup$RWELL respectively for the perpendicular tracks at $\thetaup\,=\,0^{\circ}$ and increases to  189.6~$\muup$m, 139.9~$\muup$m and  400.8~$\muup$m when the particle tracks hit the prototypes at  $\thetaup\,=\,45^{\circ}$. The degradation of the spatial resolution performance is shown in these plots with the increase of width of the residuals by more than a factor 2 for proto-I and by about a factor 4 for the reference $\muup$RWELL. The degradation of the residual width for proto-II is only about 30\% for track hits at $\thetaup\,=\,45^{\circ}$ compared to perpendicular tracks.
\subsection{Spatial resolution vs. track angle}
\label{subsec:jlab_trk_angle_scan}
\begin{figure}[!ht]
\centering
\includegraphics[width=0.45\columnwidth,trim={0pt 0mm 0pt 0mm},clip]{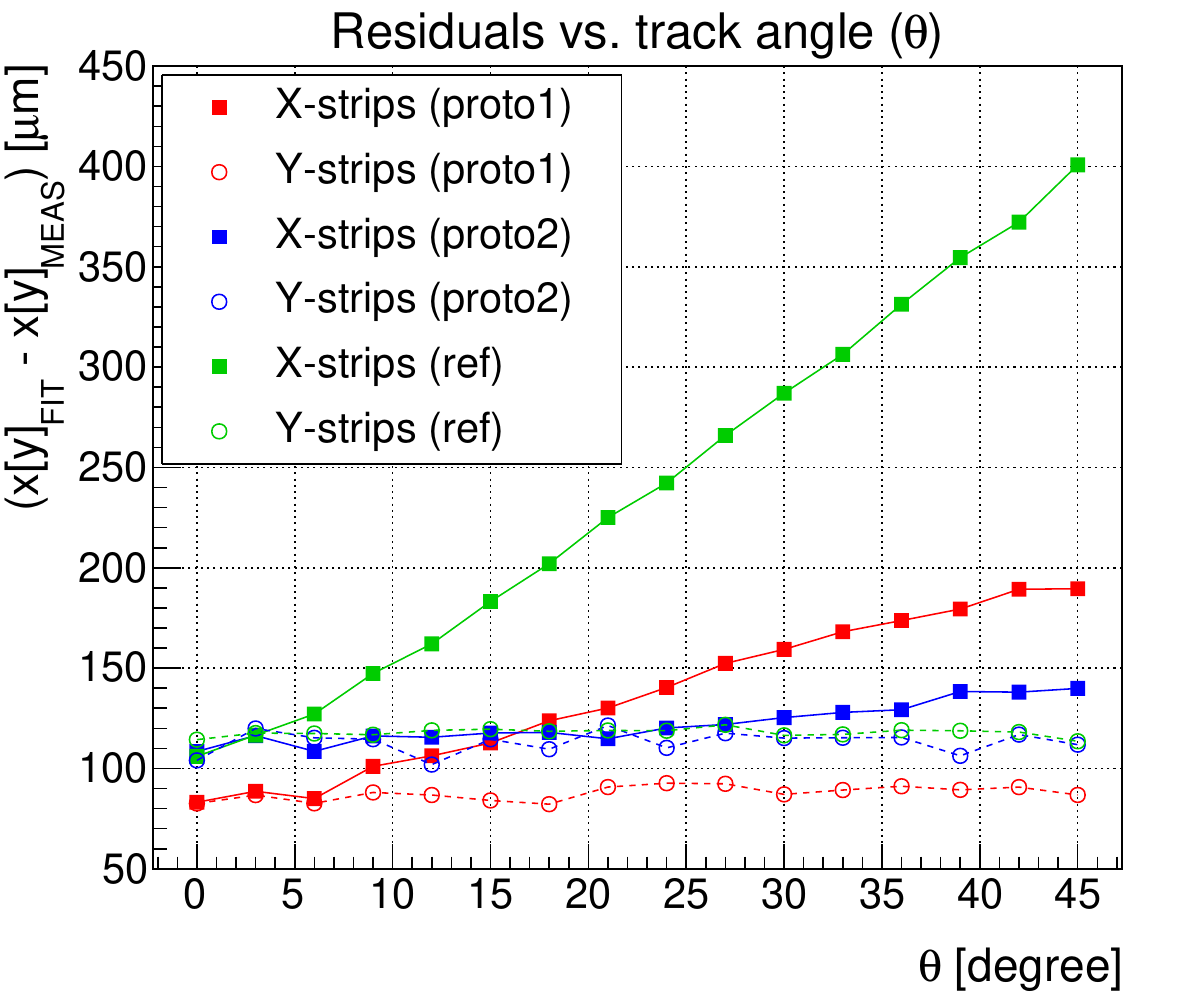}
\hspace{0.5cm}
\includegraphics[width=0.45\columnwidth,trim={0pt 0mm 0pt 0mm},clip]{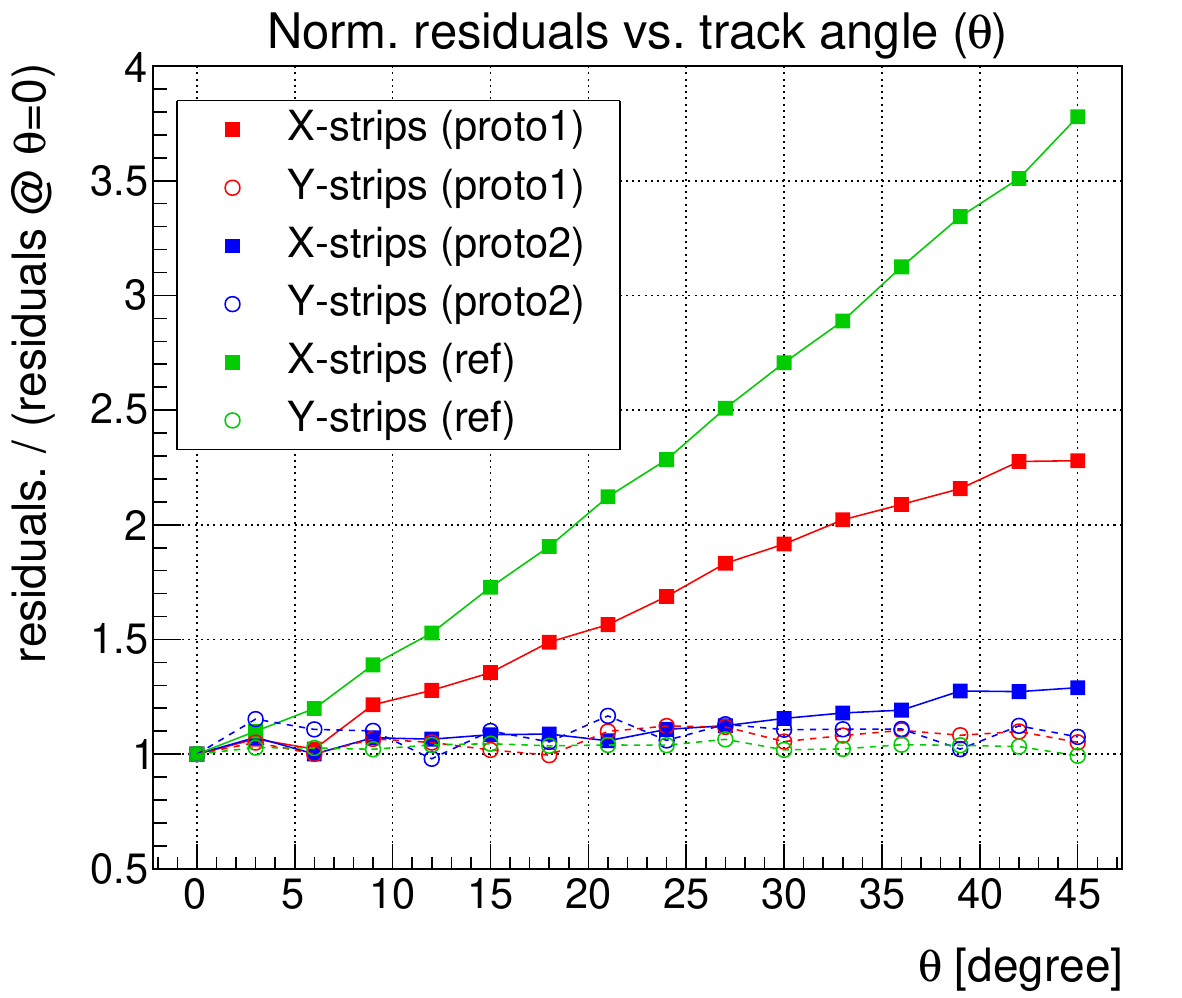}
\caption{\label{fig:jlab_res_vs_ang} (\textit{Left})  Spatial resolution  in X-strips (solid square) and in Y-strips (open circle) as a function of the track angle of the incoming particles; (\textit{right}) Spatial resolution normalized to the resolution at  0$^{\circ}$ vs. track angle.}
\end{figure}

The plots in  Fig.~\ref{fig:jlab_res_vs_ang}  show the width of the track residuals for the three prototypes as a function of angle of incoming tracks. As expected, for Y-strips (open circle) along the vertical axis where there is no rotation, the resolution is independent of the angle of the incoming tracks for all three prototypes even when the average signal collected by Y-strips increases as shown in the left plots of Fig.~\ref{fig:jlab_eff_vs_ang}. For the X-strips along the horizontal axis subjected to the plane rotation we clearly observe a linear increase of the width of the residuals as a function of the angle of the incoming tracks. As predicted, the degradation of the spatial resolution is more pronounced for the standard $\muup$RWELL prototype with 3~mm drift gap, increasing from $\sim$100~$\muup$m at $\thetaup\,=\,0^{\circ}$ to $\sim$400~$\muup$m at $\thetaup\,=\,45^{\circ}$. 
The degradation of the spatial resolution is  less severe for the 1~mm drift gap proto-I increasing from $\sim$83 $\muup$m at $\thetaup\,=\,0^{\circ}$ to $\sim$190 $\muup$m at $\thetaup\,=\,45^{\circ}$. As expected, the performance is even more impressive for the 0.5~mm gap proto-II with spatial resolution better than 140~$\muup$m at $\thetaup\,=\,45^{\circ}$, however as already discussed, it will be very challenging to achieve an efficiency better than 85\% with 0.5~mm thin gap detector configuration even with a selected gas mixture, along with difficulty in maintaining good uniformity in such small gap devices for large area detector due to mechanical constraints. 

%% file: 06_conclusion.tex
\section{Conclusion and Perspectives}
\label{sec:conclusion}
An extensive R\&D effort was conducted to explore a novel MPGD structure, the thin-gap GEM-$\muup$RWELL hybrid detector as a way to improve the spatial resolution capabilities over a wide incident angle range of the impinging particles for application of these detectors as large tracking devices in future NP and HEP experiments. The key feature of a thin-gap GEM-$\muup$RWELL detector is the reduction in the thickness of the gas volume in the drift region from  typically 3 - 4~mm in standard MPGD tracking detectors to 1~mm or less. The smaller drift gap is coupled with a hybrid MPGD amplification structure to compensate for the loss of total ionization charges in the drift volume. The hybrid configuration uses a GEM foil for electron pre-amplification followed by a $\muup$RWELL layer for a second amplification. The third key element of the thin-gap MPGD is the capacitive-sharing X-Y strip readout structure that allows excellent spatial resolution  with  reduced number of electronic  channels. \\
Two  thin-gap GEM-$\muup$RWELL prototypes, one with a 1~mm and the second with a 0.5~mm  gas thickness in the drift region  were assembled in the MPGD Lab at Jefferson Lab and tested in the proton beam at Fermilab during Summer of 2023 as part of the EIC Generic Detector R\&D FY22 program\cite{eRDG_tg_Report2023}. High voltage scans applied to the GEM and $\muup$RWELL electrode layers for gain performance measurement as well as electric field scans in the drift and transfer regions for charge collection optimization were conducted during test beam measurement to investigate the optimal operating voltage setting of the novel hybrid MPGD detectors. Test beam results indicate that 1-mm thin-gap GEM-$\muup$RWELL hybrid can reach an efficiency higher than 92\% for  both X-strips and Y-strips with standard Ar:CO$_2$ gas mixture with an optimized HV settings. The maximum efficiency reached for the 0.5~mm prototype is $\sim$80\%. The results also indicate that even with large pitch anode X-Y readout strips, a strip multiplicity of 3 can be achieved for both X-strips and Y-strips for the two prototypes, thanks to the implementation of capacitive-sharing readout structures. A strip multiplicity equal or larger than 3 guarantees a spatial resolution performance better than 100 $\muup$m. The angular scan measurement showed more than a factor two improvement in the spatial resolution  of a 1~mm  hybrid thin-gap  prototype and a factor 3 improvement with a 0.5~mm hybrid thin-gap prototype for particles hitting the detectors at   45$^{\circ}$ when compared to a standard 3~mm drift gap $\muup$RWELL. 
\begin{figure}[!ht]
\centering
\includegraphics[width=0.7\columnwidth,trim={0pt 0mm 0pt 20mm},clip]{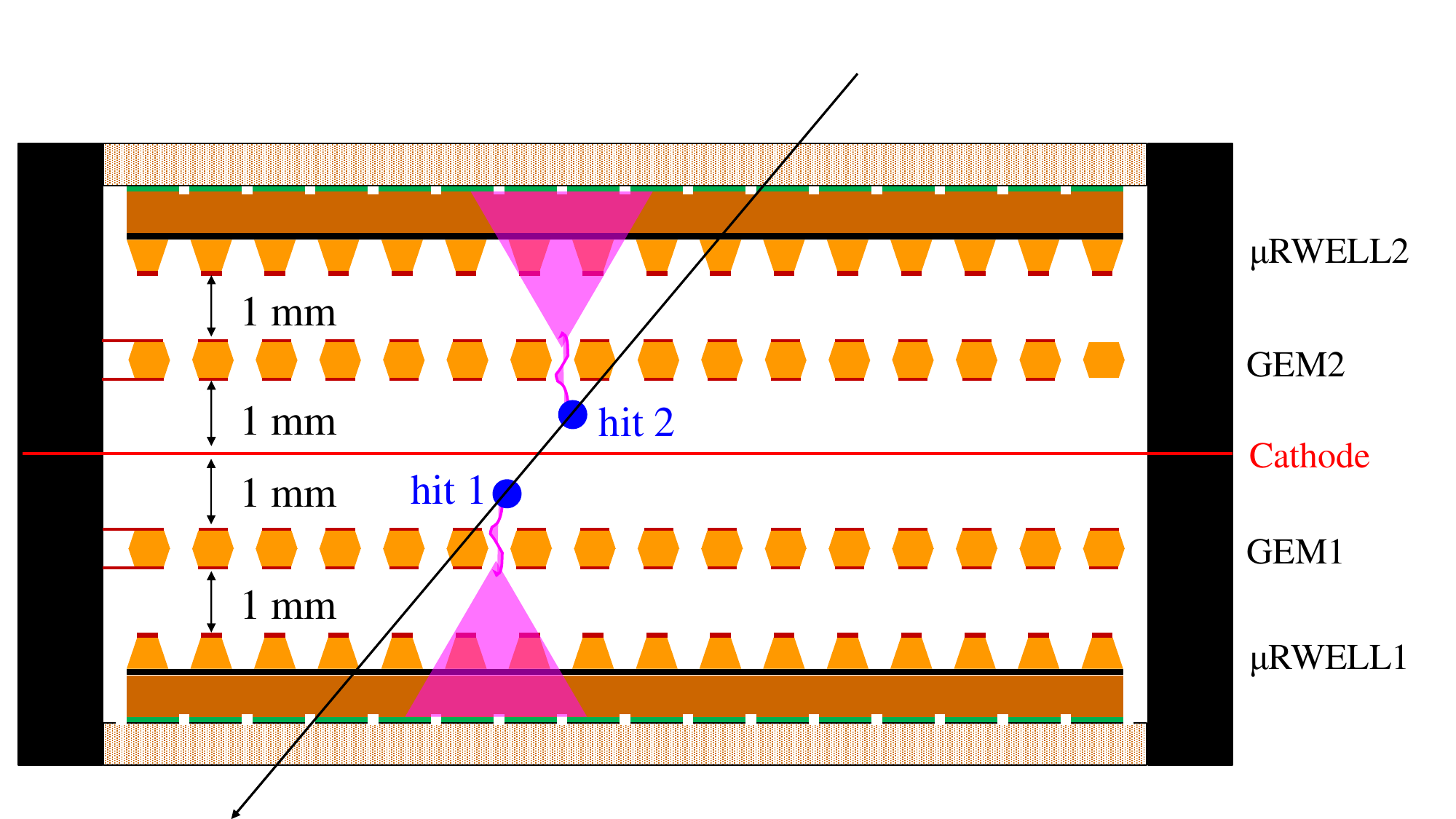}
\caption{\label{fig:double-tg-grwell} Cross sectional view of a double-sided thin-gap GEM-$\muup$RWELL detector.}
\end{figure}
For all three prototypes, a spatial resolution of $\sim$60 $\muup$m could be reached for perpendicular tracks with higher gain~\cite{capaSh_urwell2022} than the maximum gain tested in beam at the FTBF.  The preliminary results of the thin-gap  GEM-$\muup$RWELL hybrid prototypes indicate that the technology is a viable option for high performance, large area tracking systems for future NP and HEP experiments. R\&D efforts are ongoing for the full validation of the concept of thin-gap  GEM-$\muup$RWELL hybrid detector on large area trackers.  A  large (1700~mm~$\times$~330~mm) GEM-$\muup$RWELL hybrid chamber is under construction as the project engineering design test article~\cite{epic_urwell-bot} for the large $\muup$RWELL Barrel Outer Tracker ($\muup$RWELL-BOT) in the central tracker of the ePIC experiment for the Electron Ion Collider (EIC). \\
The concept of double (or 2-in-1) GEM-$\muup$RWELL hybrid detector is also been explored as part of the EIC Generic Detector R\&D program~\cite{eRDG_tg_Report2023}. The conceptual design of a double GEM-$\muup$RWELL hybrid detector is shown in Fig.~\ref{fig:double-tg-grwell}, where two GEM-$\muup$RWELL hybrid amplifications coupled each with capacitive-sharing 2D-strip readout planes are stacked face-to-face to share a common  drift cathode and a single gas volume. In this configuration, the particle crossing the detector, has a $\ge$~92\% probability to produce a hit in either amplification region, leading  to $\geq$99\%  probability for a minimum of one hit in the detector and $\geq$84\% for a double hit. A double-sided thin-gap GEM-$\muup$RWELL prototype~\ref{fig:double-tg-grwell} with an active area of 300 mm $\times$ 300 mm is under development to demonstrate that a efficiency better than 99\% and an improved spatial resolution performance are achievable with standard Ar:CO$_2$ gas mixture.
\section{Acknowledgments}
\label{sect_acknowledgements}
The authors thank the staff at the Fermilab Test Beam Facility (FTBF), including Mandy Rominsky, Eugene Schmidt, Joe Pastika and Todd Nebel, for their support and expertise during the beam test campaign. This material is based upon work supported by the U.S. Department of Energy, Office of Science, Office of Nuclear Physics under contracts DE- AC05-06OR23177. 
We also acknowledge support from the DOE funded and Jefferson Lab managed EIC-related generic detector R\&D program. Finally I would like to thank Ms. Nicole Gnanvo for the invaluable support.